\documentclass{emulateapj}
\usepackage{amsmath}
\usepackage{epsfig}
\usepackage{multirow}

\newcommand{\be}   {\begin{equation}}
\newcommand{\bmax} {{b_{\rm max}}}
\newcommand{\bmin} {{b_{\rm min}}}
\newcommand{\ee}   {\end{equation}}
\newcommand{\tew}  {\overline{T}}
\newcommand{\zew}  {\overline{Z}}

\begin{document}

\title{SMAUG: A NEW TECHNIQUE FOR THE DEPROJECTION OF GALAXY CLUSTERS}
\author{
	Fabio Pizzolato\altaffilmark{1, 2}, 
	Silvano Molendi\altaffilmark{1},
	Simona Ghizzardi\altaffilmark{1},  
	Sabrina De~Grandi\altaffilmark {3}
}
\altaffiltext{1}{IASF/CNR Sez. "G. Occhialini" Via E. Bassini 15/A 20133 
Milano - Italy; fabio@mi.iasf.cnr.it.}
\altaffiltext{2}{Dipartimento di Scienze, Universit\`a dell'Insubria, 
Via Valleggio 11, 22100, Como - Italy.}
\altaffiltext{3}{Osservatorio Astronomico di Brera, via Bianchi 46,
22055 Merate (LC), Italy.}


\begin{abstract}
This paper presents a new technique for reconstructing the spatial 
distributions of hydrogen, temperature  and metal abundance of a 
galaxy cluster. These quantities are worked out from the X-ray 
spectrum, modeled starting from few  analytical functions describing their 
spatial distributions.  These functions depend upon some parameters,
determined by fitting the model to the observed spectrum.
We have implemented this technique  as a new 
model in the {\it XSPEC} software analysis package.
We describe  the details  of the method, and apply it to work out the structure
of the cluster A1795. We  combine the observation of three satellites, 
exploiting the high spatial  resolution of {\it Chandra} for the cluster core,  the 
wide  collecting area of  {\it XMM-Newton} for the intermediate regions and
the large field of view of  {\it Beppo-SAX} for the outer regions.
We also make a threefold  test on the validity
and precision of our method by  $i)$ comparing its  results with those from
a geometrical deprojection, $ii)$ examining the spectral residuals at
different radii of the cluster and $iii)$ reprojecting the unfolded 
profiles and comparing them directly to the measured quantities. 
Our analytical method yields the parameters defining the spatial functions
directly from the spectra.
Their  explicit knowledge  allows a straightforward  derivation of other
indirect physical quantities like  the gravitating mass,   as well as 
a fast and easy estimate of  the profiles uncertainties.
\end{abstract}




\section{INTRODUCTION}
\label{sec:intro}

A deep insight into  the early life of our Universe  is provided by
the largest and youngest virialized structures: clusters of galaxies.
A valuable tool to fathom the physics and the structure of these objects  is their X-ray emission.
This stems  from the hot ($1-10$~keV) and optically thin 
plasma  (also known as intra-cluster medium, or ICM) 
which makes up some $10-20\%$ of their total mass.
This rarefied plasma is so hot because it has to develop a  
pressure strong enough to maintain itself in quasi-hydrostatic equilibrium within  
the gravitational well of the dark matter. The ICM is transparent and allows to 
see deeply, down to the very core of the cluster. This transparency has 
also a drawback: photons born at different depths
mix up along the same lines of sight,  and a direct use of the spectral information
to investigate in three dimensions the physical conditions of  a cluster is impossible.
The first step is necessarily a {\it deprojection}  of the  X-ray  spectrum, which   
provides the  deconvolved
three-dimensional profiles of density,  temperature and  metal abundance. 
From these fundamental direct quantities others may follow, like
the distribution of heavy elements or  the  cluster  gravitating mass.

This paper describes a novel deprojection technique   we have implemented  as a 
new model in the  {\it XSPEC} software package for the analysis of X-ray spectra.
The fundamentals  of the theory are recalled in  Section \ref{sec:cloud}.  
In Section \ref{sec:ana} we apply them to build  up our deprojection model.
In Section \ref{sec:a1795} we make a specific example by deprojecting 
the nearby cluster A1795. A comparison is also made between our results and those from 
another  deprojection technique. In Section \ref{sec:reject} the methods to test the solidity of
our deprojection are reviewed.  Finally, in Section \ref{sec:gmass} we 
show how the result of our analytical deprojection may help to  derive  other 
physical variables of cosmological interest.

\section{THE SPECTRUM OF AN OPTICALLY THIN SPHERE OF PLASMA}
\label{sec:cloud}

In this Section we briefly review  the fundamental formulae for the flux  from the diffuse medium 
of a cluster of galaxies. This is approximated as  a  spherically symmetric  cloud of optically 
thin plasma located at  redshift $z$. The observed photon flux from a source subtending the
solid angle $\Omega$ is
\be
\label{eq:flux}
{\cal F}_E =  \int_\Omega d\Omega\: I_E/E,
\ee
where $I_E$ is the specific brightness in the observer's rest frame, 
expressed in terms of the spectral energy $E$.
The specific brightness is obtained by 
integrating the spectral emissivity $j_{E'}$ along an optical path across 
the source. If this passes at a distance $b$ from the  emission center, 
the hypothesis of spherical symmetry yields
\be
\label{eq:emi}
I'_{E'} = \int dl\: j_{E'} = 
2 \int_b^\infty{dr\frac{r\: j_{E'}(r)}{\sqrt{r^2-b^2}}},
\ee
where all primed quantities refer to the cluster rest frame, so $E=E'/(1 + z)$ is the redshifted 
spectral energy.  
The quantity $I_{E}/E^3$ is invariant along the path of a light ray
(see e.g. Rybicki \& Lightman 1979), thus the emitted and observed specific brightnesses
are related by
$I'_{E'}/E'^3 = I_{E}/E^3$. Taking this into account, Equation~(\ref{eq:emi}) 
can be substituted into (\ref{eq:flux}), where the integration 
over the solid angle can be 
usefully rewritten in terms of  the projected size of the source.
If the emitting region is located at the angular distance $D_A$ and is
seen as an annulus bounded  by the projected rims 
$\bmin$ and $\bmax$, Equation~(\ref{eq:flux}) becomes
\be 
{\cal F}_E  = \frac{4\, \pi}{D_A^2\,(1+z)^2}
\int_{\bmin}^{\bmax} db\,b \int_b^\infty dr \frac{r}
{\sqrt{r^2-b^2}}\:\epsilon_{E(1+z)}(r),
\label{eq:double}
\ee
where  $\epsilon_E =j_E/E$ is the photon emissivity. The spectral  
emissivity is $j_{E}=n_H\, n_e\, \Lambda_E(T,\, Z)$, where $n_e$  and $n_H$  
are the  electron and hydrogen  densities: the temperature  $T$  and the metal 
abundance $Z$ enter  $j_E$  only through the spectral cooling 
function $\Lambda_E$.
The occurrence of a double integral
in the evaluation of a spectrum is a computationally  tough problem.
The possibility of writing a reasonably  efficient {\it XSPEC} code to 
model the  spectrum of  a cluster lies in the possibility of reducing 
this double  integral to a single integral. Under the hypothesis of spherical 
symmetry this can be done, 
and the reader is referred to Appendix A for the technical 
details. The resulting formula 
\begin{gather}
\label{eq:spec}
{\cal F}_E = \frac{4\:\pi}{D_A^2\,(1+z)^2} \int_{\bmin}^\infty{\hskip -3mm dr\, {\cal K}(r; \bmin,\bmax)\, \epsilon_{E(1+z)}(r)}\\
\label{eq:kern}
{\cal K}(r; \bmin,\bmax) = 	
	\begin{cases}
	r\,\sqrt{r^2-\bmin^2}            & \text{if $r < \bmax$} \\
	r\left(\sqrt{r^2-\bmin^2}\,-\sqrt{r^2-\bmax^2}\right)
	& \text{if  $\bmax < r$},
	\end{cases}
\end{gather}
is the basis of our deprojection method. 

\section{THE PROBLEM OF 	DEPROJECTION}
\label{sec:ana}

\subsection{The geometrical deprojection}
\label{sec:gd}

The problem of recovering the 3-D shape of $T$, $n_H$ and $Z$ is usually tackled with 
the methods of geometrical  deprojection. Our term  ``geometrical''  encompasses  a variety of
non-parametric techniques to extract   3-D quantities from their
corresponding 2-D projected counterparts. This Subsection provides a summary of 
the main  geometrical deprojection techniques, as well as some references for further readings on 
the subject.

The method of deprojection in the study of galaxy clusters was introduced by
Fabian et al. (1981), and Kriss, Cioffi \& Canizares (1983) in their pioneering works on the analysis of
{\it Einstein}  data. The spectral resolution of the instruments {\sl HRI} and {\sl IPC} 
on board {\it Einstein} was too poor for a direct deprojection of the spectra, 
and these authors unfolded the cluster starting from its  surface brightness. In the hypothesis 
of spherical symmetry, the  image of the cluster is divided into $N$ rings of radii 
$r_1 < r_2 < \cdots < r_N$ centered on  the emission peak. 
The 3-D emission is supposed to come from $N$ homogeneous spherical shells bounded
by the same radii. The observed surface brightness of the $j$-th projected ring (of area $A_j$) is
\begin{equation*}
S_j = \frac{\sum_{i=j}^N C_i  \; V_{i, j} }{A_j},
\end{equation*}
where  $V_{i, j}$ is the volume of the fraction of the $i$-th spherical shell subtended by the $j$-th 
projected  ring. This matrix equation can be solved for  the spatial count emission  rates $C_i$ of each
shell. In turn, these quantities are converted to photon emission rates with the effective area of
the instrument. Since the emission rate  depends both on temperature and density, 
in the absence of further spectral information this degeneracy hinders  any  determination 
of  these quantities. In oder to determine separately the 3-D density and  temperature profiles, 
Fabian (1981) was forced to assume an average metallicity  of the cluster, and to suppose the gas 
to be in  hydrostatic equilibrium within an {\it assumed} gravitational potential.

When spectral information became available,  the hypothesis of hydrostatic equilibrium 
could be avoided, thus allowing a substantial extension of this deprojection technique.
Nulsen \& B\"ohringer (1995)   adopted a variant of  Fabian's  method to analyze  
 {\it ROSAT} spectra:  it was applied to the count  rate of each  energy channel, 
instead  of the integrated spectrum. This  procedure  yielded the 
pulse height spectrum of each spherical shell, which was converted to photons and fitted to 
an emission  model to obtain the  3-D values of temperature and hydrogen density. 

David et al. (2001) --~see also Allen, Ettori \& Fabian (2001) and  Buote (2000)~--
modeled the flux from any  annulus as  a weighted  sum of the
emissivities of the spherical shells.  Each shell  was supposed to be characterized 
by its  own uniform temperature, density and metal abundance, considered as the 
parameters of the fit, determined by fitting simultaneously the annular spectra.

Ettori et al.  (2002) devised a matrix method to deproject spectra. 
The spectrum of each ring is first analyzed to
determine its  emission weighted projected temperature,  metal abundance, emission integral 
and luminosity
\begin{gather*}
T_{\rm ring}(j)   = \frac{\sum_{i = j}^N \epsilon_i V_{i,j}\: T_i} {\sum_{i = j}^N \epsilon_i V_{i,j}} \\
Z_{\rm ring}(j)   = \frac{\sum_{i = j}^N \epsilon_i V_{i,j}\: Z_i} {\sum_{i = j}^N \epsilon_i V_{i,j}} \\
EI_{\rm ring}(j)  =  \sum_{i = j}^N  V_{i,j}\: n_H \, n_e\\
L_{\rm ring}(j)  =  \sum_{i = j}^N  V_{i,j}\:  \epsilon_i,
\end{gather*}
where  $n_e \simeq 1.21 n_H$ is the electron density, $V_{i, j}$ has the same meaning as before, 
and $\epsilon_i$ is the emissivity (in erg cm$^{-3}$ s$^{-1}$)  of the $i$-th shell.
Since the volumes are known, the previous matrix equations can be solved to yield the 3-D
quantities appearing on their right-hand sides.

\subsection{The analytical deprojection}

Whenever a spectral information is available,  the deprojection techniques 
described in the  last Section do not require the introduction of  any  ``theoretical bias'' in the 3-D 
profiles. Nevertheless, the  profiles recovered 
are often rather  wiggly. Thus, for many applications they need to be regularized,
causing the loss of some information on the fine-grained patterns of  the X-ray emitting gas.   
Another common requirement   of all geometrical techniques is the lack of gaps between
nearby rings, since they  would bar from knowing  the exact number  of  photon  counts to 
subtract and   proceed  inwards.

In this paper we introduce an alternative approach to deprojection  in which this restriction 
is  not necessary,  and where no {\it a posteriori} regularization is required.
The only condition our method does not relax is the hypothesis of spherical symmetry, under
which  all the relevant deprojection formulae discussed  in  this paper have been deduced.
Our method looses some fine-grained information (exactly as the geometrical
techniques do), the difference being that here the loss occurs {\it before} the deprojection. 
In our {\it analytical deprojection}
we model the X-ray emission from the cluster, i.e. we {\it do}  introduce a ``theoretical bias'' by giving
to the 3-D distributions of $T$, $n_H$ and $Z$  known functional forms,
 dependent  on  some parameters $a_1,a_2,\cdots a_N$.  With   Equation~(\ref{eq:spec}) we
calculate the model  spectrum   ${\cal F}_E[a_1,a_2, \cdots a_N]$  as a function of these parameters,
fit it to  the observed  spectra and determine the best-fit values of $a_1,a_2,\cdots a_N$, 
fixing the  deprojected 3-D  distributions of  $T$, $n_H$ and $Z$. 
These profiles do not need any 
regularization, simply because --strictly speaking--ours is {\it not} a ``deprojection'', but a 
procedure to determine the parameters of  smooth  3-D functions
which better match the 2-D observed profiles. One might question the strength
of the  bias introduced by forcing the 3-D distributions to be described by a
functional form with a (necessarily) small number of parameters. This important issue
shall be discussed in Section \ref{sec:reject}.

We have implemented our deprojection  technique as a new model for the {\it XSPEC} analysis 
software  package and have christened it {\sl smaug} after the dragon
of J.R.R. Tolkien's tale {\it The Hobbit}, but the name also  stands for 
{\bf S}pectral {\bf M}odeling {\bf A}nd  {\bf U}nfolding of {\bf G}alaxy clusters.
Given a first guess of the fit parameters  $a_1,a_2,\cdots a_N$, {\it XSPEC} 
calls {\sl smaug} and  calculates for it 
the spectral cooling  function  $\Lambda_E(T, Z)$. Then
{\sl smaug} sums up the local contributions to the spectrum
by evaluating the right-hand side of Equation~(\ref{eq:spec}),
and passes back the result to {\it XSPEC}. Finally, {\it XSPEC}  
compares the model spectra to the  observed spectra and works out an 
improved estimate of 
$a_1,a_2, \cdots a_N$. The procedure is iterated until the values of 
the fit parameters are stable.

The evaluation of the  {\sl smaug} model requires the calculation of 
the integral
(\ref{eq:spec}), deduced from the double integral (\ref{eq:double}) 
under the hypothesis of spherical symmetry. This reduction is
necessary, since  double integrals are  not easy to evaluate 
numerically. Besides, the integrand function contains  the 
spectrum $\epsilon_E(r)$, i.e. a function not only of $r$, but also of the
energy  $E$. Without the reduction of the double integral,
the evaluation of a model spectrum  would be exceedingly time-consuming 
from the computational viewpoint.
Even the  numerical calculation  of the single integral (\ref{eq:spec}) 
requires some care in order to keep as light as possible the computational 
effort without loosing accuracy. We have confined all technical considerations 
about this important topic to Appendix B, and the interested reader is
referred there.  

The model is fed with a collection of spectra extracted in
annuli centered about the X-ray emission peak. 
Having specified the cosmology, the user is 
prompted for a series of parameters. The first are the redshift $z$ 
of the source, the rims $\bmin$ and $\bmax$ (in arcmin) of
each extraction ring, as well as the  cutoff radius 
(in Mpc)  and the  number of mesh points for the calculation of 
the integral (\ref{eq:spec}).
Two  further fixed parameters specify whether the spectrum is to be 
calculated or interpolated form existing tables, and which type of plasma 
emission code  has to be used: the choice is among 
{\tt  Raymond-Smith}, {\tt Mekal}, {\tt Meka}, or {\tt APEC}.
The  variable fit parameters of the model define the
distributions of hydrogen density, electron temperature and metal abundance
within the cluster.

The hydrogen density profile is modeled with  a double-beta function,
fixed by six  parameters:
\be
\label{eq:nh}	
n_H(r) = n_0\left[f\, n_c(r)\: + \: (1 -f) \,n_g(r)\right],
\ee
where
\be
n_{c,g}(r) = \left[1 + \left(\frac{r}{r_{c, g}}\right)^2\right]^{-\beta_{c, g}}.
\ee
The first beta-component refers to the ICM, and the second  to the dominant galaxy 
(if any)  lying  at the center of the cluster.  The variable $n_0$ is the central hydrogen density, 
and $f$  is the fraction of $n_0$ relative to the ICM. 
The rationale for the choice (\ref{eq:nh}) is that double beta-models 
have been widely and successfully applied to fit the density distributions 
in several clusters (see e.g. Mohr, Mathiesen \& Evrard 1999).

The next group of eight parameters defines the temperature profile
\be
\label{eq:kt}
\left\{
\begin{array}{l}	
	T(r) = k(r)\left[ T_0 + T_1 l(r) \right]\\
	k(r) = [1 + (r/r_{\rm tail})^2]^{-\omega}\\
	l(r) = \frac{2}{\pi}\frac{\arctan\left[(r/r_{\rm iso})^\varkappa \right]}{1 + (r_{\rm cool}/r)^\xi}. 
\end{array}
\right.
\ee
Despite its appalling appearance, this formula has a simple  astrophysical  rationale.
There are three regions with characteristic radii $r_{\rm iso}$,
$r_{\rm cool}$ and $r_{\rm tail}$,  the hierarchy of which is expected to be
$r_{\rm iso} < r_{\rm cool} < r_{\rm tail}$. 
At small radii, i.e. for $r \lesssim r_{\rm iso}$, the plasma is nearly 
isothermal, being  $T \sim T_0$. This core flattening of $T$ has been
recently observed with {\it Chandra} in some clusters 
(A2199: Johnstone et al. 2002; A1835: Schmidt, Allen \& Fabian 2001;
A1795: Ettori et al. 2002).
At intermediate radii $ r_{\rm iso} \lesssim  r  \lesssim r_{\rm cool}$
the temperature  $T \sim T_0 + T_1 l(r)$ increases with radius, as 
observed in practically all relaxed clusters harboring a cD galaxy at their 
center (see e.g. De~Grandi \& Molendi 2002). At  radii larger than 
$r_{\rm cool}$ the temperature flattens, and finally for $r \gtrsim  r_{\rm tail}$ 
it starts to decrease, behaving  essentially as a beta
function, in the same fashion as the hydrogen  density.
In the outskirts of  the cluster  $T \propto r^{-2\omega}$ and 
$n_H \propto r^{-2\beta_c}$, thus the equation of state of the 
ICM is expected to be a  polytrope $T \propto n_H^{\omega/\beta_c}$.
The decrease of $T$ at large radii was first observed by Markevitch et al. 
(1998)  with  {\it ASCA} in a sample of 30 clusters, and more recently 
by De~Grandi \& Molendi (2002) with {\it Beppo-SAX}.

It is worth to remark  that an apparently complicated profile like (\ref{eq:kt})
is actually highly modular: by  suitably fixing some parameters, the
user may impose the temperature profile to be only increasing, only decreasing,
or even isothermal. A function  like (\ref{eq:kt}) is able to encompass a wide
range  of qualitative behaviors of the spatial temperature profile.

The metal abundance  profile is described by the three parameters function
\be
\label{eq:ab}
Z(r) = Z_0 \: \left[ 1 + \left(\frac{r}{r_Z}\right)^2\right]^{-\zeta}.
\ee
This choice is motivated by the observation of a radial decrease in the metal
concentration in practically all clusters  (De~Grandi \& Molendi 2001), and
by more recent {\it Chandra} and {\it XMM-Newton} observations  showing  a 
flattening of the metallicity in the core of M87/Virgo (Molendi \&
Gastaldello 2001) and other clusters (e.g. A496: Tamura et al. 2001).
For the time being we have not included in the functional form of $Z(r)$ 
the decrease in metallicity observed in the core of Perseus
(Schmidt, Fabian \& Sanders 2002) and of Centaurus (Sanders \& Fabian 2002). 
If called for by future  observation, it will be easy to modify
the profile (\ref{eq:ab}) to properly describe  this behavior.

It is known that {\it XSPEC} rescales any additive model by multiplying 
it for a 
normalization constant $\cal N$ (Arnaud \& Dorman 2000), thus the model 
parameter $n_0$  in Equation~(\ref{eq:nh}) is degenerate with 
respect to $\cal N$. Before starting a fit,  $n_0$ must  
be fixed  to unity.   The normalization ${\cal N}$  imposed to {\sl smaug} by 
{\it XSPEC} has the meaning of the  square of the  central hydrogen 
density  in cgs units, so  $ {\cal N}^{1/2} = n_H(r=0)\;{\rm cm}^{-3}.$

As a final remark, we would like to stress that our model is extremely
flexible: the  code has been written with a modular structure, allowing an easy 
change  of the functional forms  of $n_H$, $T$ or $Z$, whenever new and better 
X-ray observations will require it. 
For instance, to modify  the temperature  profile (\ref{eq:kt}) , 
the user  has to 1) replace the code fragment returning the temperature with a more
suitable one, 2) edit the parameters input section in the  main program, and finally 
3) edit the {\tt lmodel.dat}  {\it ASCII} file telling {\it XSPEC} what are the model parameters
(Arnaud \& Dorman 2000).
We hope that the good   flexibility might render {\sl smaug}   (and its possible future
implementations) a useful tool  for the whole community.

\section{A SPECIFIC EXAMPLE: THE DEPROJECTION OF ABELL 1795}
\label{sec:a1795}

\subsection {Data preparation}
\label{sec:a1795:data}
In this Section we describe how our deprojection code works in practice.  
We deproject the spectrum of A1795, a  nearby ($z=0.0631$) and relatively 
relaxed cool-core cluster (Ettori et al. 2002).
For our analysis we have taken a flat Universe with
$\Omega = 1.0$, $\Omega_\Lambda= 0$, and 
$H_0 = 50$~$\rm km \,s^{-1}\,Mpc^{-1}$.
In order to cover the whole cluster, we have used  the observations of 
three satellites, namely {\it Chandra},  {\it XMM-Newton} and 
{\it Beppo-SAX}. 
Their views are complementary:  {\it Chandra}'s sharp angular resolution 
(below 1 arcsec, corresponding to $1.7$~kpc at the distance of A1795) 
is optimal  for observing  the core of the cluster.
{\it XMM-Newton} has a broader PSF, but thanks to  its wider field of view
and large collecting area is particularly well-suited for observing the intermediate regions. 
In the outskirts, where {\it XMM-Newton}'s view is contaminated by a   relatively high 
background  and angular resolution is less critical,  {\it Beppo-SAX} is a better choice.
It is worth noting that the composite view from these satellites covers about two decades of the
cluster extension, ranging from about 10~kpc to 1.5~Mpc. 

The  $19.5$~ks  {\sl ACIS-S} {\it Chandra} observation
has been taken with a focal plane temperature of --$110$ degrees. 
The analysis has been performed with version V2.2  of the CIAO  
software and with the calibration data base CALDB 2.12. 
The details on the preparation  and analysis of the {\it XMM-Newton} data are 
the same as in Molendi \&  Pizzolato (2001), the only difference being that
here the  new version V5.2 of the SAS software has been used. 
The preparation of the data from {\it Beppo-SAX} is described in 
De~Grandi \& Molendi (2001).

In order to check the results of analytical deprojection, we have also  undertaken the analysis 
with Ettori's matrix  geometrical method. In the remainder of the paper we
shall often address Ettori's method simply as ``geometrical deprojection''. The
reader should remind, however, that this is only one of the possible
geometrical deprojection techniques. 

\subsection{The geometrical deprojection}
\label{sec:a1795:geom}

We have selected  5 spectra from {\it Chandra}, 4 from {\it XMM-Newton}, 4 
from {\it Beppo-SAX}, and have accumulated  them in concentric rings 
centered about  the X-ray emission peak.
They have been analyzed using  version  11.0.1 of {\it XSPEC}. 
A spectral analysis is common to the geometrical and the analytical 
deprojection.
In the first case the analysis provides the 2-D emission-weighted quantities, 
still to be deconvolved.  In the second, the spectral analysis directly yields 
the parameters of the 3-D distributions of the relevant physical quantities.

Both spectral analyses include a {\sl wabs} multiplicative component for 
the Galactic  absorption ($1.21\times 10^{20}~{\rm cm^{-2}}$).
A variable {\sl constant} factor describes a moderate 
($\lesssim 10\%$) cross-normalization  uncertainty among 
{\it Chandra}, {\it XMM-Newton}  and {\it Beppo-SAX}. 
Since the spectra are not simultaneously fitted
in the case of geometrical deprojection, the value of these 
cross-normalization factors must be provided. 
We have taken their values from the result of the (simultaneous) fit of the 
spectra  performed with the {\sl smaug} model described in the next Section.
A further fixed geometrical {\sl constant} factor has been
introduced to account for the  excision of  the portion  of the
{\it Beppo-SAX}  detectors shaded  by the entrance windows.
Since geometrical deprojection proceeds inwards, and necessarily only a finite number  
of rings can be used,  the contamination from the  emission of the cluster outskirts 
must be considered. We have modeled this cutoff  effect as described by  
McLaughlin (1999).
Table \ref{tab:summ} summarizes the radii of the extraction regions of the
spectra, their energy ranges, the geometrical corrections and the 
cross-normalization factors for the 13  datasets employed.

The spectral analysis preliminary to the geometrical deprojection has been
performed with a model {\sl wabs*constant*constant*mekal}. The  {\sl mekal} 
component yields the  emission integrals and the (projected) emission weighted 
temperature and  metallicity of any ring, from which the relevant 3-D profiles
can be extracted  as  described at the end of Section \ref{sec:gd}.
The error on these  deprojected profiles is estimated with a Monte Carlo 
technique. We build a sample of randomly distributed values around the
mean, assuming Gaussian errors. From this sample a  statistical collection of 
1000 2-D profiles is obtained and deprojected: the
standard deviation of the  deprojected sample
provides the required estimate of the error on the 3-D profiles. 
The outcomes  relative to  geometrical deprojection are shown by the 
marks and error bars in Figures  \ref{fig:nh}, \ref{fig:kt} and \ref{fig:ab}. These Figures also plot the 
analytically deprojected results  described  in the following Subsection.

\subsection{The analytical  deprojection}
\label{sec:a1795:anal}

In the language of {\it XSPEC},  our analytical model is 
{\sl wabs*constant*constant*smaug}. The meaning of the first three 
components has already been discussed: {\sl smaug}
provides the parameters for the 3-D distributions of hydrogen density, 
temperature and metal abundance. 

The analytical model  has 15 free parameters: two 
cross-normalization  constants for
the satellites and 13 parameters for {\sl smaug}. We have calculated  the
X-ray emission from A1795 with the {\tt Mekal} emission code. 
The cutoff radius and the 
number of mesh points for the evaluation  of the integral (\ref{eq:spec}) 
have been 
fixed respectively  to 2~Mpc and 15. We have checked {\it a posteriori} that 
the best-fit values of the parameters thus determined are not 
sensitive to these choices. In particular,  changing the cutoff radius to 3~Mpc 
and rescaling the number of mesh points, the values of 
the best-fit parameters do not change appreciably, indicating that there are no important
numerical  effects associated with the evaluation of the integral (\ref{eq:spec}).

Not all the parameters available in {\sl smaug} have been employed for the 
analysis of A1795:  like all parametric models,  {\sl smaug} calls for some {\it a posteriori}
fine tuning of the parameters,  suggested by the results of the fitting procedure.
The exponent $\varkappa$ of the temperature gradient at
intermediate radii during the fit  systematically  pegs 
against its upper hard limit $\varkappa = 10$, where we have chosen  to freeze it.
For this cluster the temperature distribution  does not depend much  on the cooling radius 
$r_{\rm cool}$, which is rather ill-determined by the fit procedure.
Therefore, we have decided to set it equal to the 
radius of the cluster density  beta component $r_c$. 
The polytropic index $\gamma$ of the plasma in the outskirts is poorly constrained by 
the {\it Beppo-SAX} datasets, whose statistics is rather poor with respect to {\it Chandra} and 
(above all)  {\it XMM-Newton}. For this reason we have fixed its value, taking it from 
the statistical analysis  by De~Grandi \& Molendi (2002) of the temperature profiles of a sample of  
{\it Beppo-SAX}  clusters. These authors find an average  $\gamma = 1.2$ for cool-core
clusters. Accordingly, in our analysis we have  fixed  $\omega/\beta_c = \gamma-1 = 0.2$.

The metal profile is a rapidly declining function of the radius, and therefore the parameter $Z_0$ 
is rather uncertain.  When left free the radius $r_Z$ assumes the unreliably small value 
of 1~kpc, comparable to the angular size of {\it Chandra}'s PSF  at the distance of A1795. 
We have therefore frozen $r_Z$ to  8.24~kpc, i.e. one-half of the innermost {\it Chandra}'s radial bin.
The parameter $Z_0$ assumes the physical meaning of the metallicity of the central bin.

The deprojected profiles of density, temperature and metal abundance are plotted in 
Figures~\ref{fig:nh}, \ref{fig:kt} and \ref{fig:ab}. The best-fit values of the free parameters are 
shown in Table \ref{tab:best} with their 
$1$-$\sigma$ errors. The  reduced chi-square of the best fit is 
$\chi^2 / {\rm dof} = 3652 / 3031 \simeq 1.20$, 
but its associated probability is negligible, only $3.2 \times 10^{-14}$. This  value 
is clearly underestimated: indeed, our analysis uses the spectral data
from  three different satellites, and thus a degree of cross-calibration error is  unavoidable. 
We have partly kept notice of this by introducing two variable constants 
in our spectral analysis,  but this is not good enough, since the spectral  fit procedure tends 
to interpret the systematic  cross-calibration differences as statistical errors, 
thus lowering the  probability of the fit.  We shall return in Section~\ref{sec:reject}
to the problem of testing an analytical  spectral model in presence of this effect.

\begin{figure}
\epsscale{1.00}
\includegraphics[angle=-90, width=85mm]{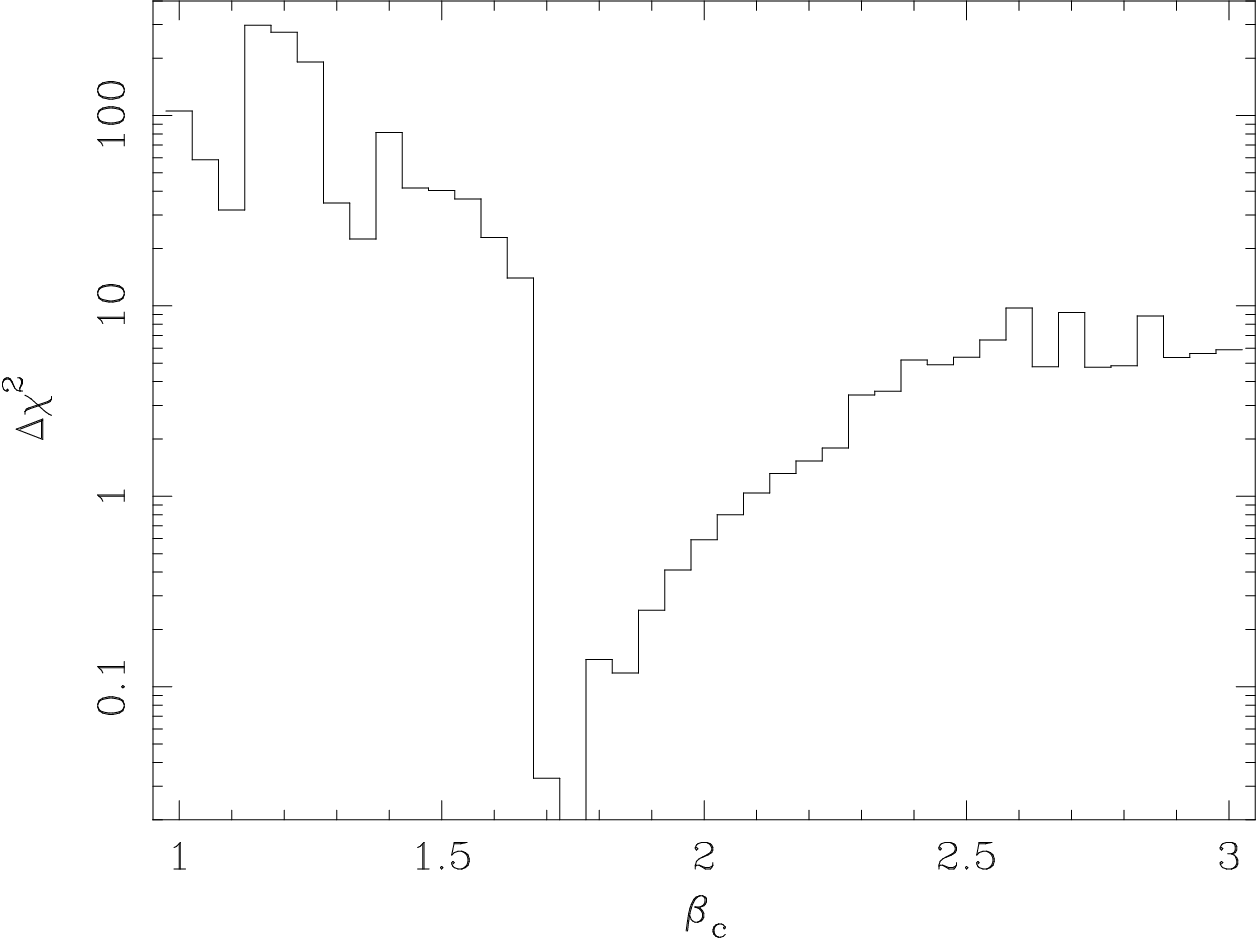}
\includegraphics[angle=-90, width=85mm]{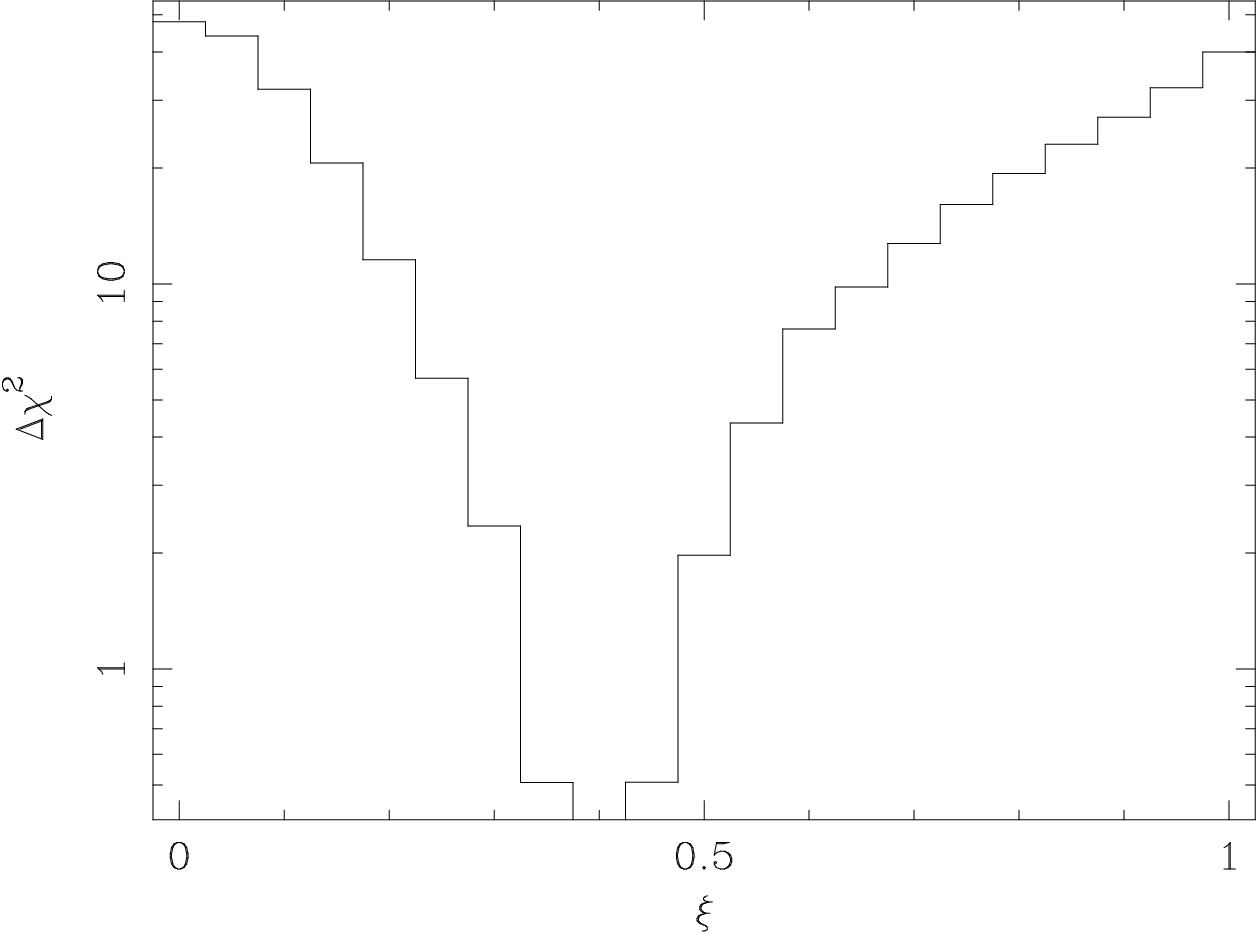}
\figcaption{The $\Delta\chi^2$ on two fit parameters, obtained by running the {\it XSPEC}
command {\it steppar}. Left panel: the $\Delta\chi^2$ on the  exponent $\beta_c$ of the hydrogen 
 cluster component. Despite the presence of secondary minima, the best-fit value of this 
parameter is robust. Right panel: the same plot for the exponent $\xi$ of the temperature profile
shows a much smoother behavior.
\label{fig:deltachi2}}
\end{figure}

Like all models characterized by many fitting parameters, also  ours may be affected by the 
problem of secondary minima.  A clear  example is the  rather complex  $\Delta \chi^2$ pattern 
of  the exponent $\beta_c$ relative to   the hydrogen density, plotted in the left panel 
of Figure  \ref{fig:deltachi2}.  The absolute minimum is well defined, but the plot is asymmetric, with
several secondary minima. Complex behaviors are not always the rule:  a much more
well-behaved  $\Delta \chi^2$ is shown in the right  panel of the same Figure, and refers to  
the temperature exponent $\xi$. These examples demonstrate that it is always a good practice 
to check the results against  the possibility of secondary minima, before trusting blindly a 
deprojected profile.

\subsection{Inspection of the best fitting spectra}
\label{sec:a1795:spec}
 
We now turn our attention to the problem of checking  the quality of our  model for A1795.
We have already remarked that  in our case the  $\chi^2$ test is  of relatively little use, being 
affected by the systematic cross-calibration effects arising from the use of data sets from
different satellites. For this reason, in this and the next two Subsections  we explore  
alternative routes.

\begin{figure}
\epsscale{1.00}
\includegraphics[angle=-90, width=5cm]{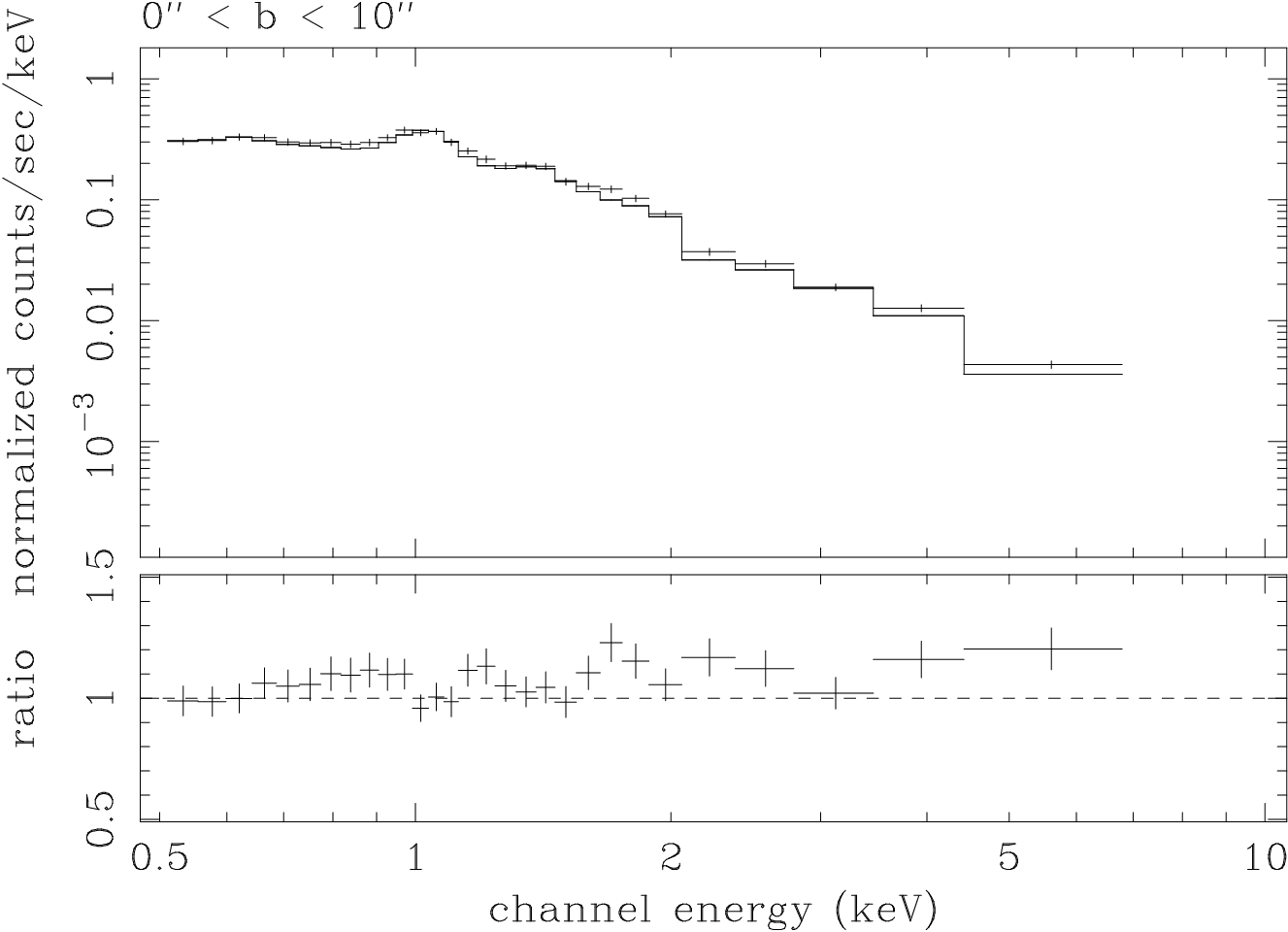}
\includegraphics[angle=-90, width=5cm]{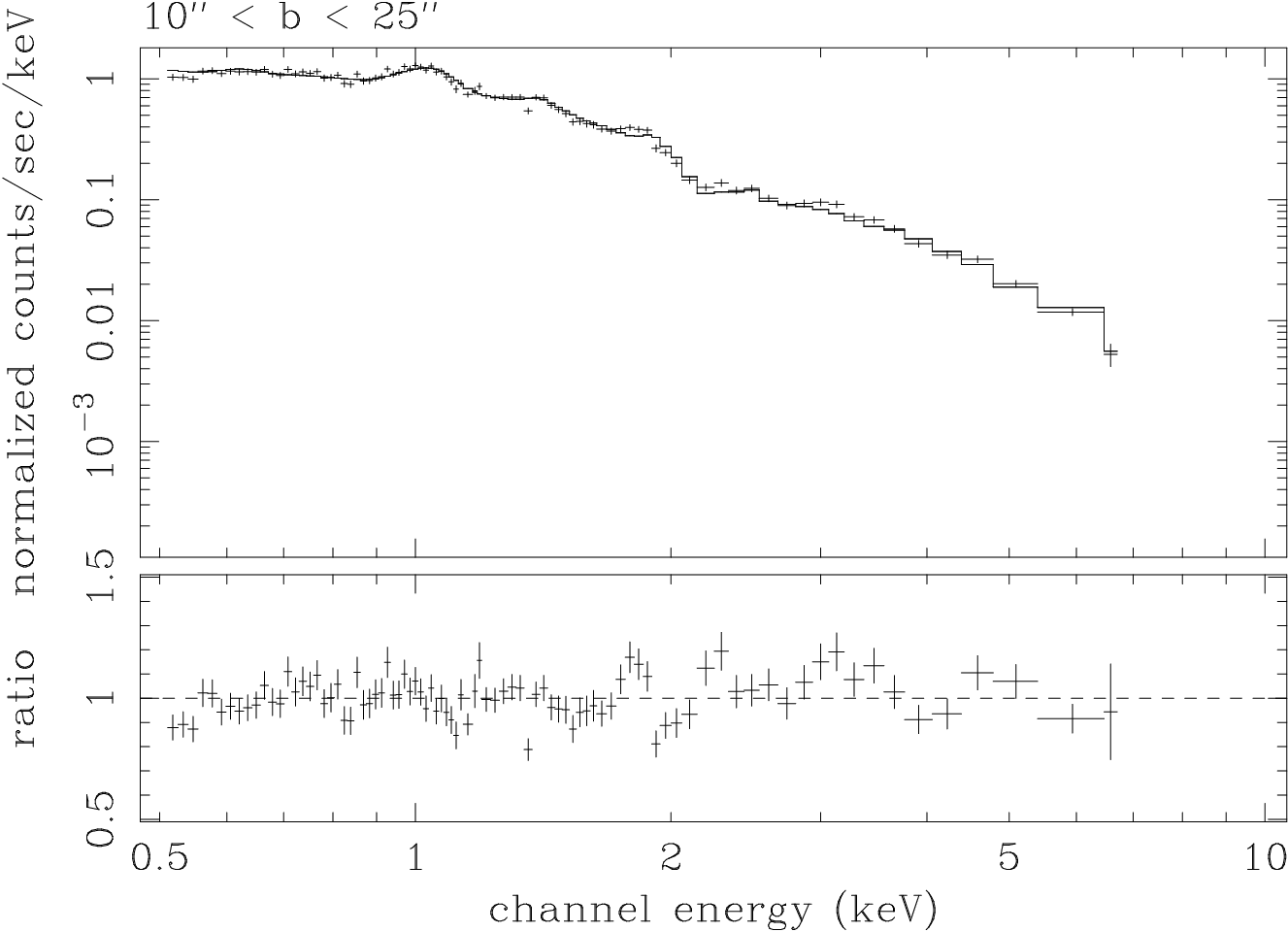}
\includegraphics[angle=-90, width=5cm]{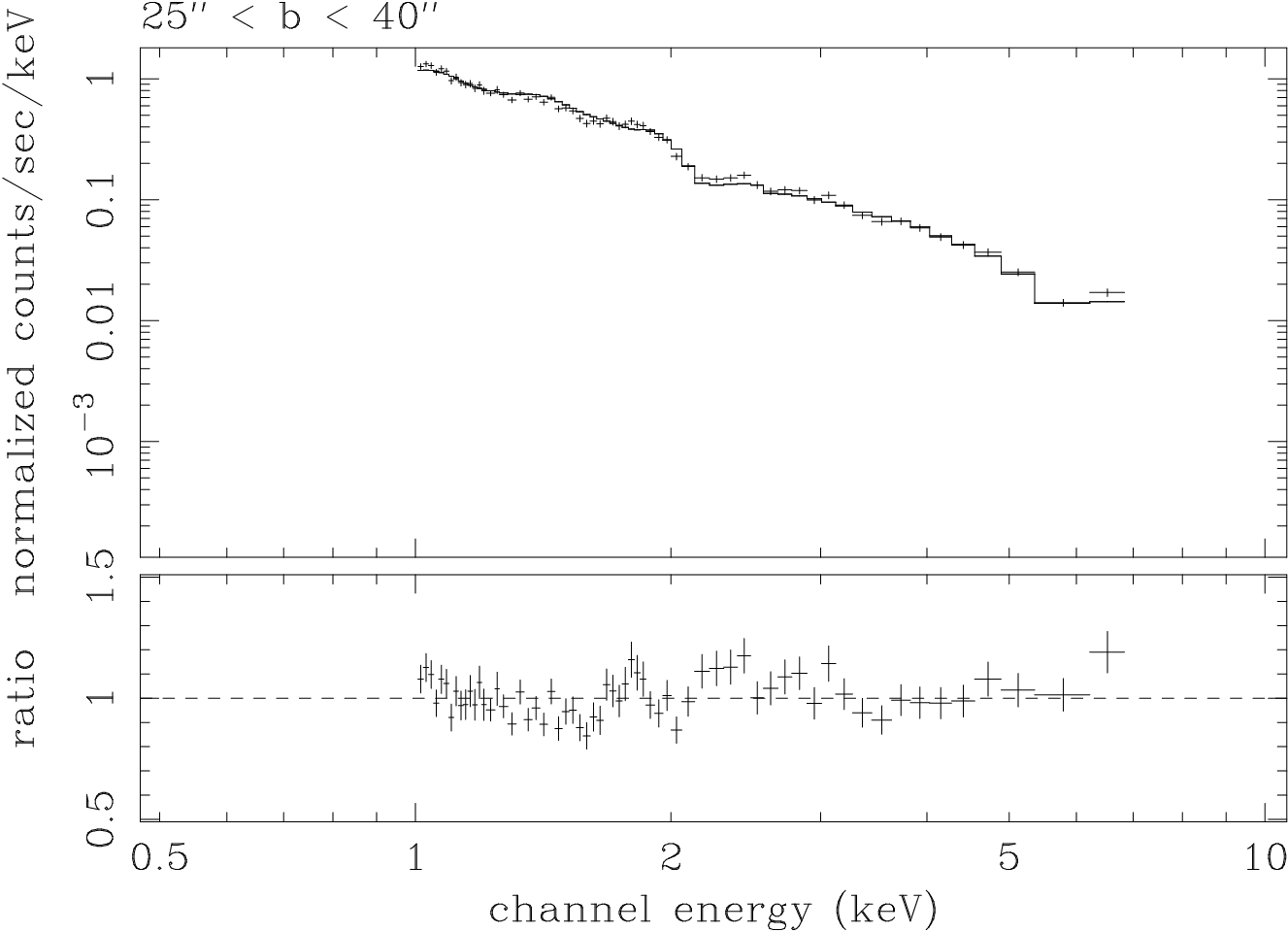}
\includegraphics[angle=-90, width=5cm]{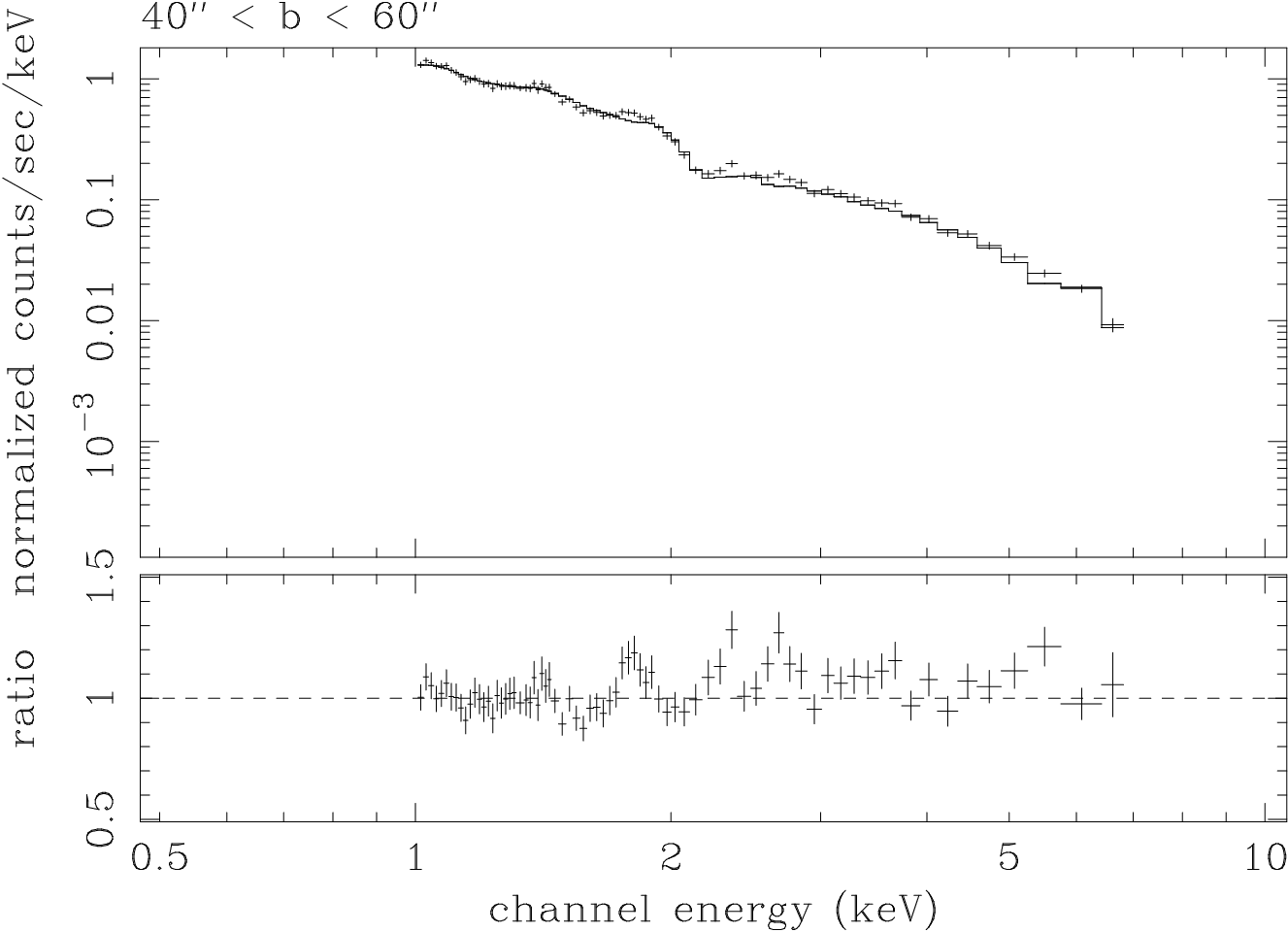}
\includegraphics[angle=-90, width=5cm]{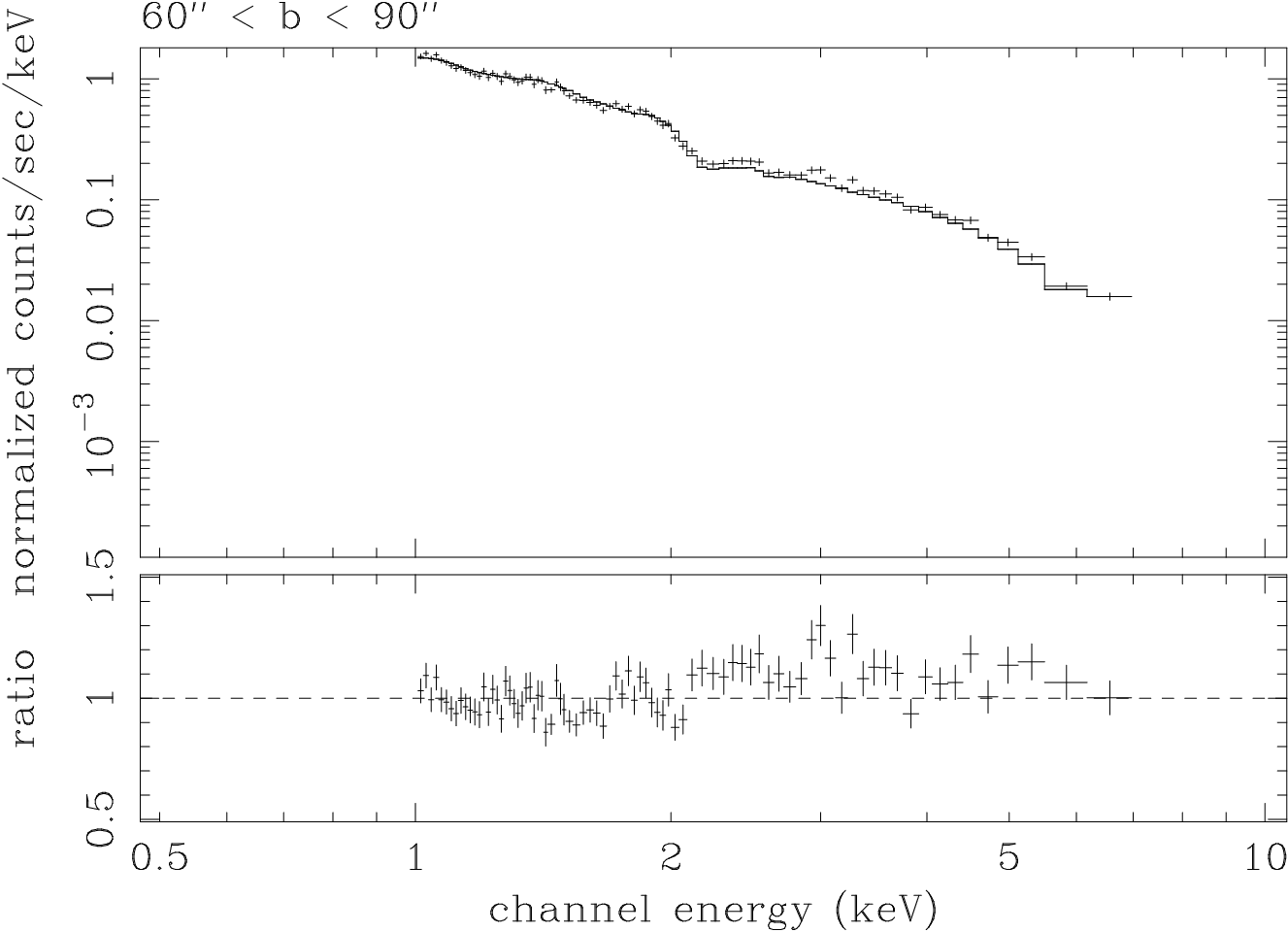}
\includegraphics[angle=-90, width=5cm]{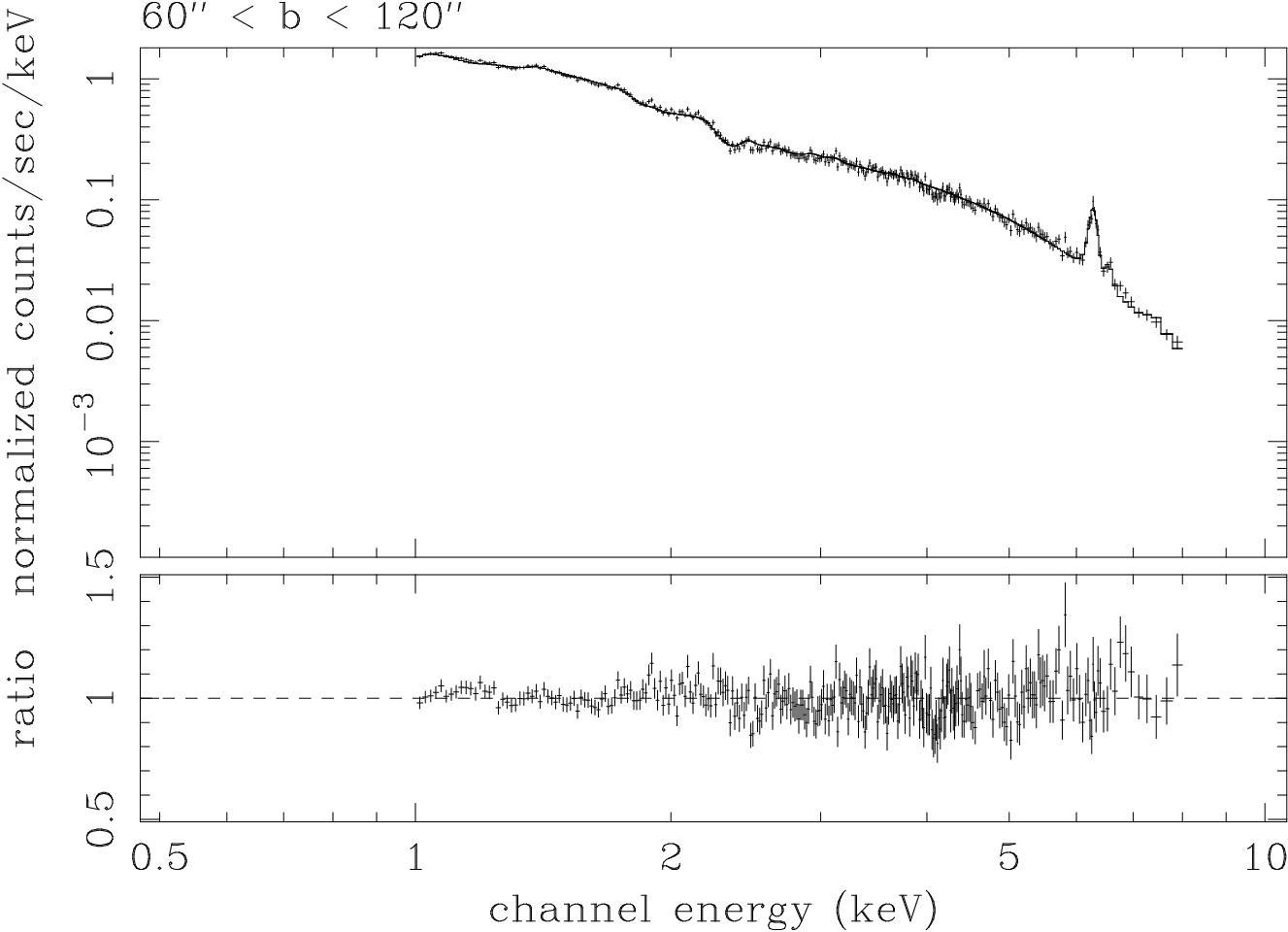}
\includegraphics[angle=-90, width=5cm]{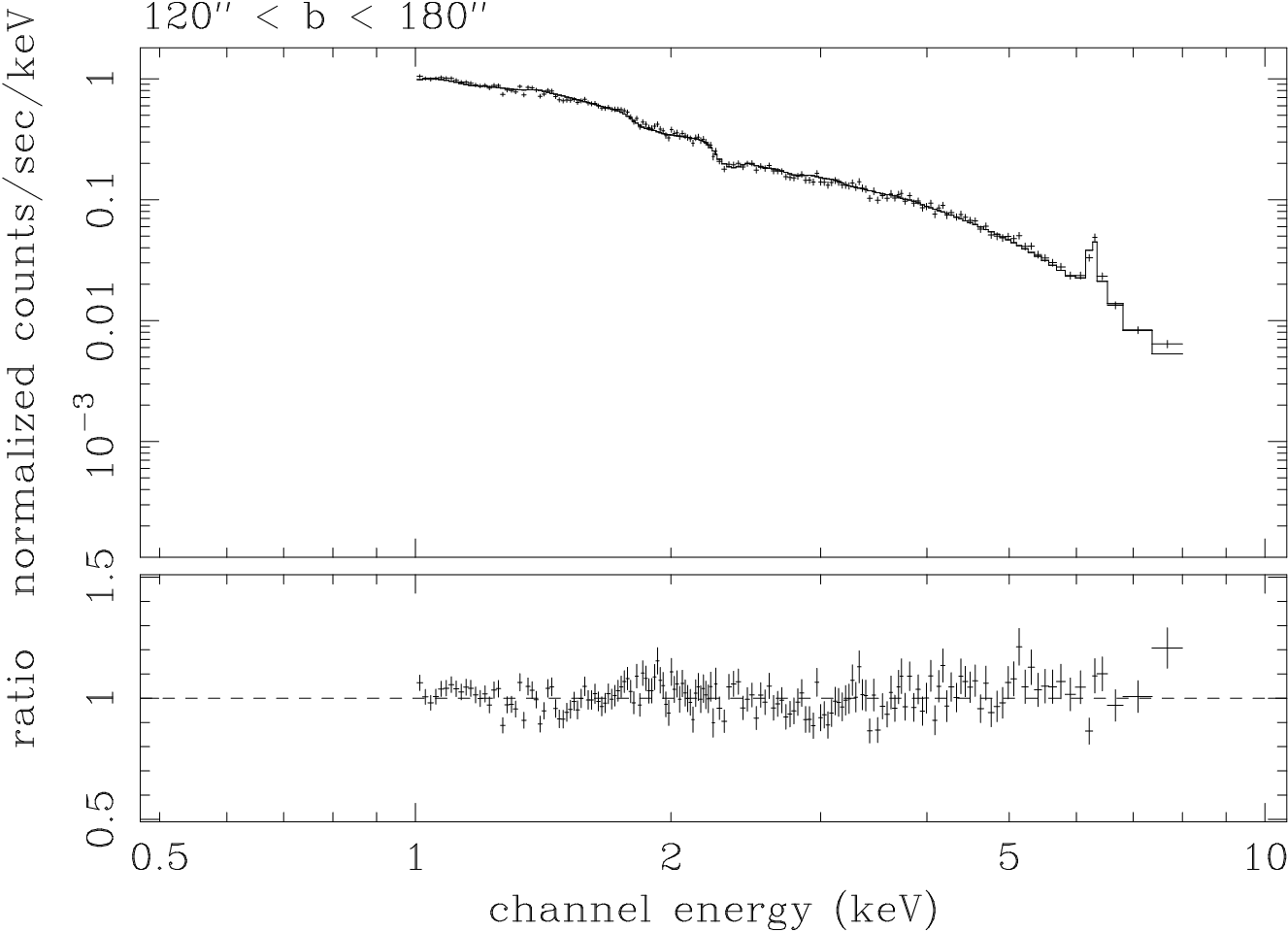}
\includegraphics[angle=-90, width=5cm]{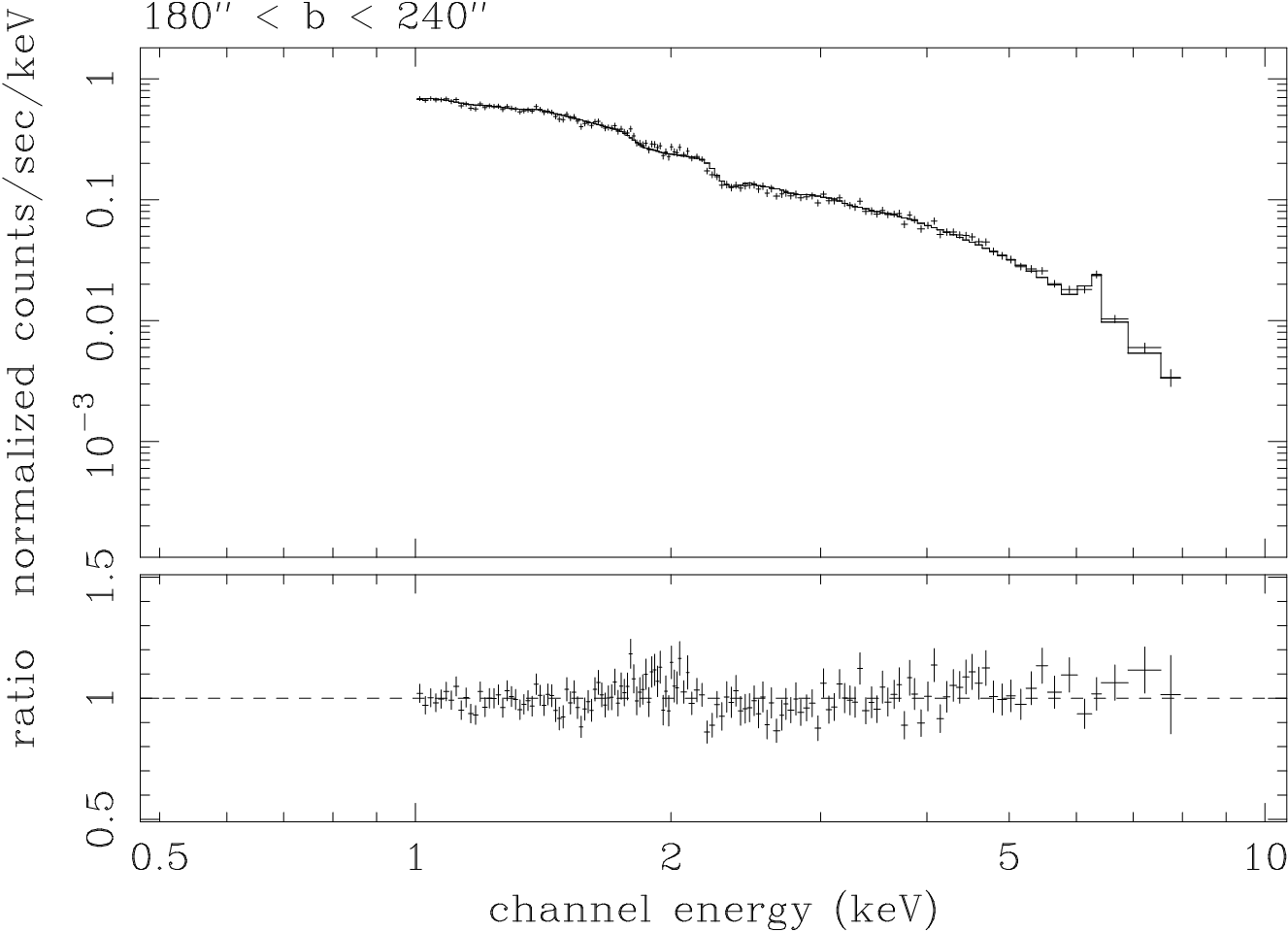}
\includegraphics[angle=-90, width=5cm]{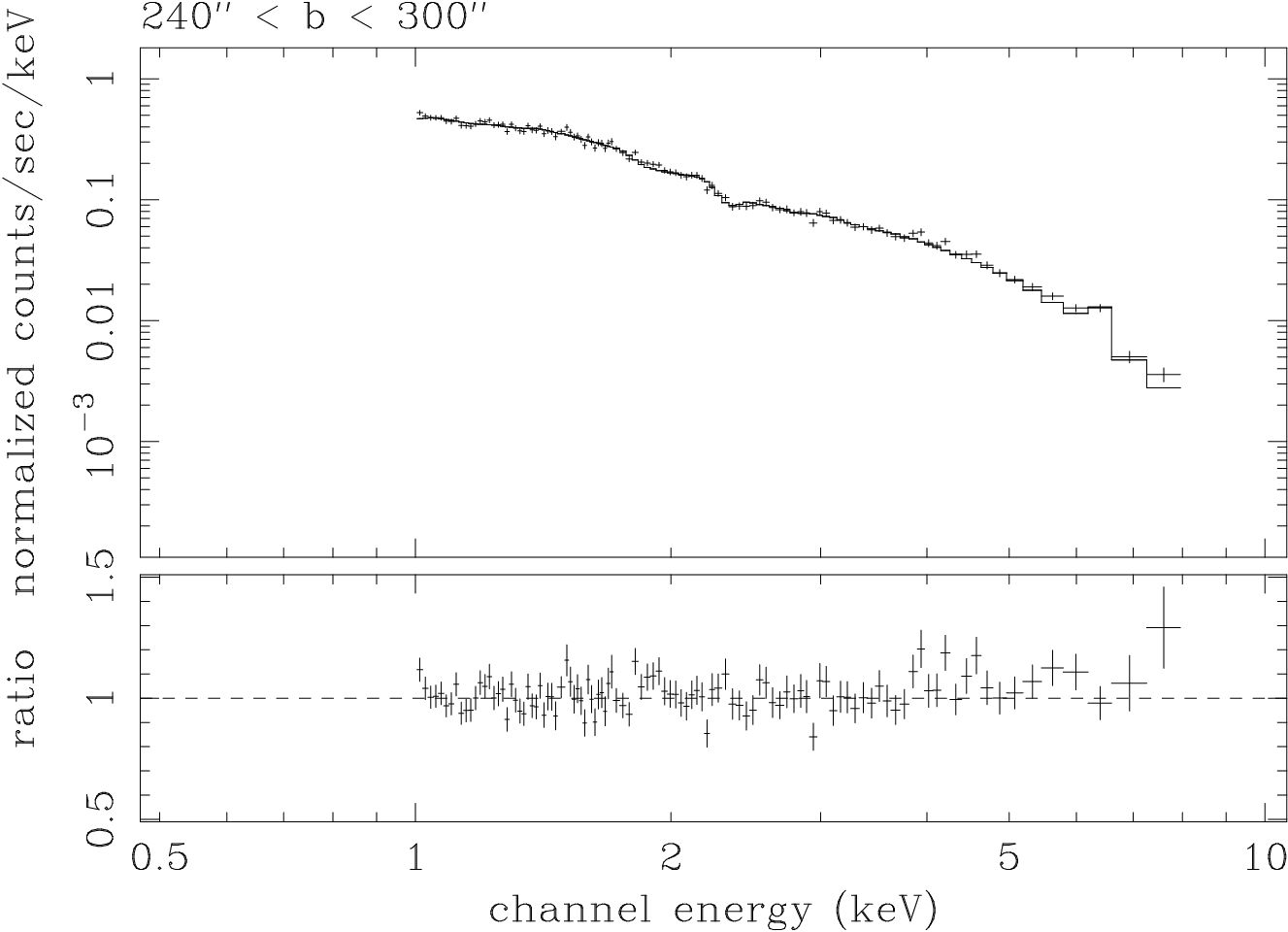}
\includegraphics[angle=-90, width=5cm]{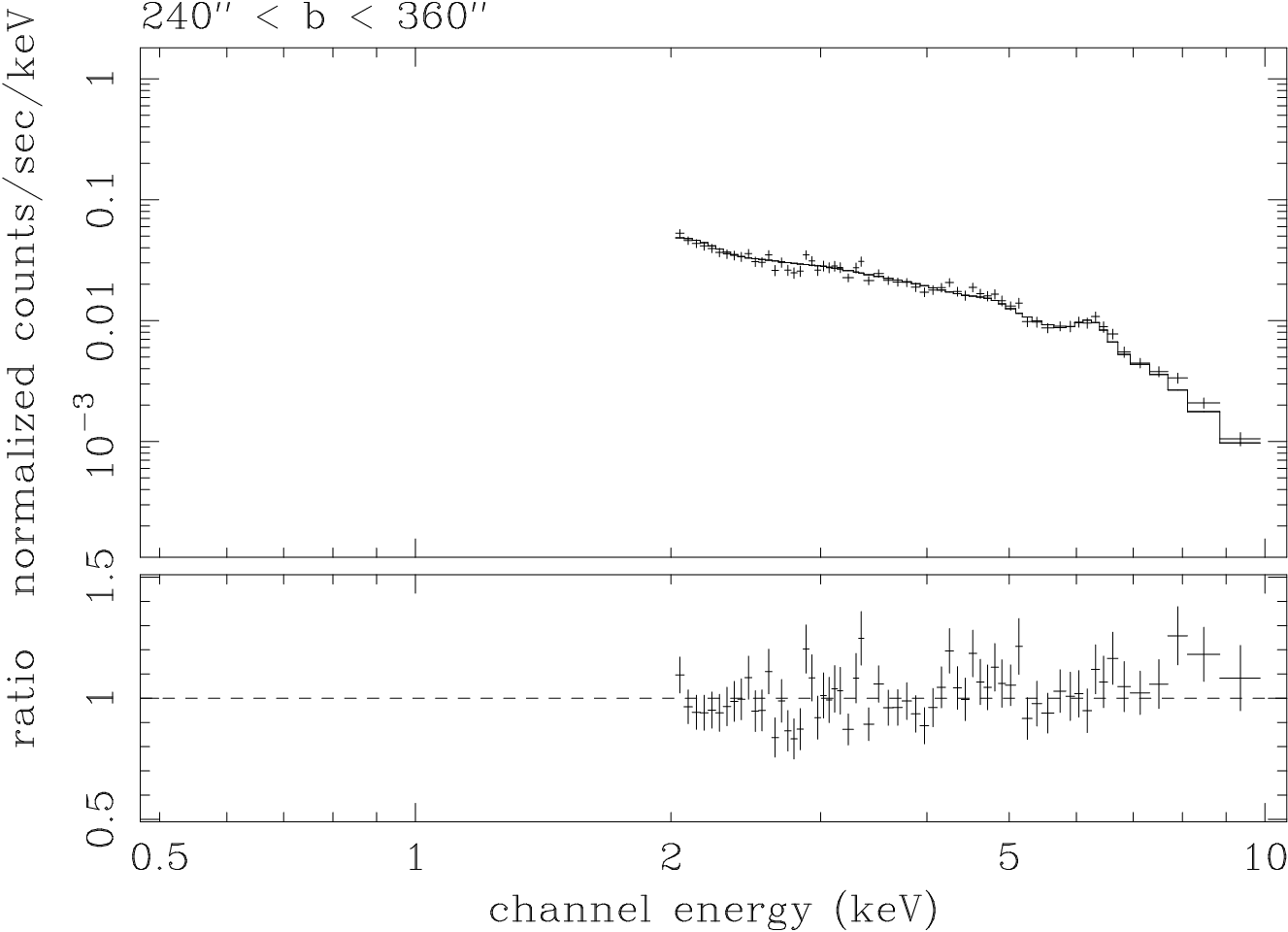}
\includegraphics[angle=-90, width=5cm]{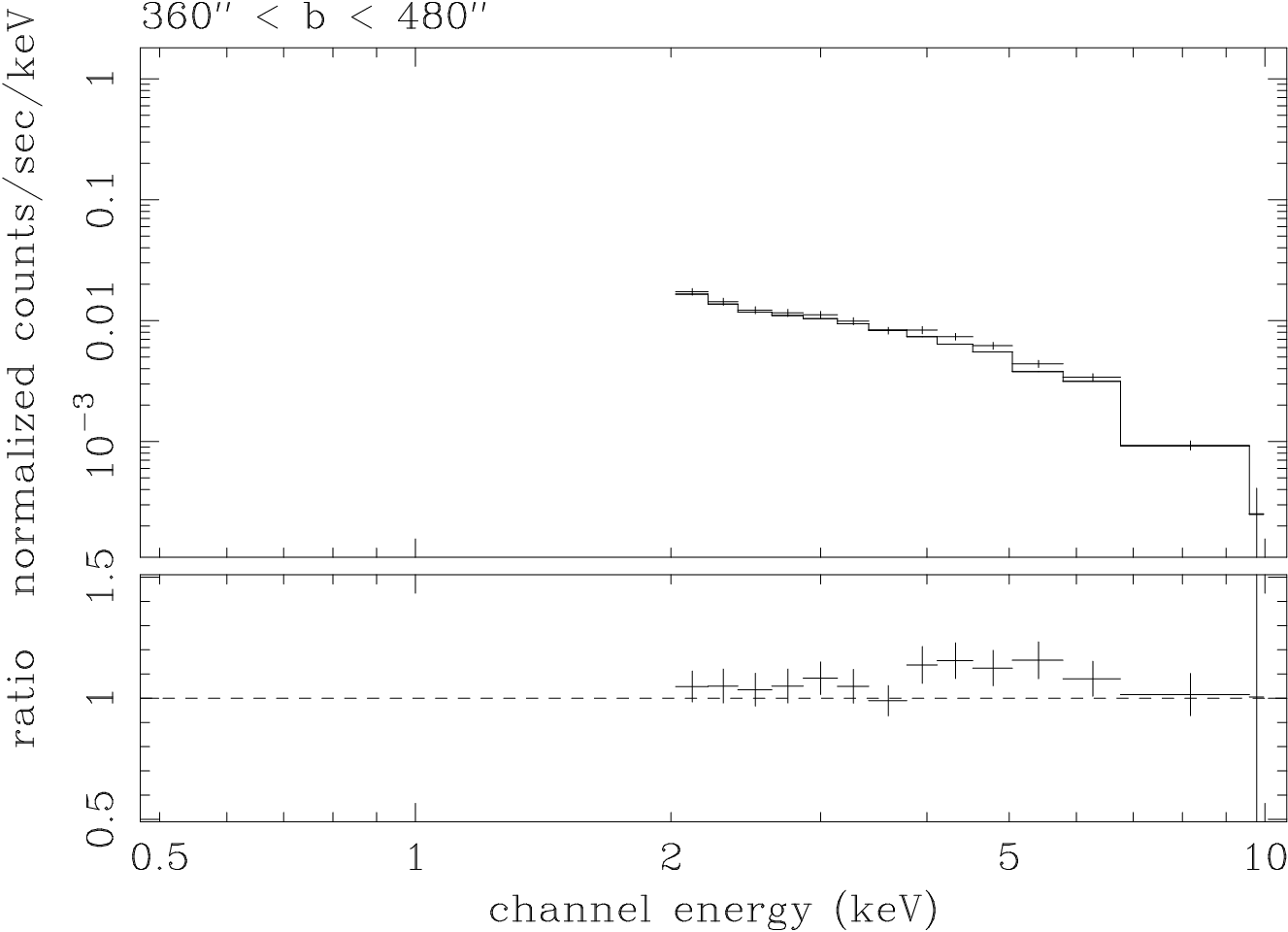}
\includegraphics[angle=-90, width=5cm]{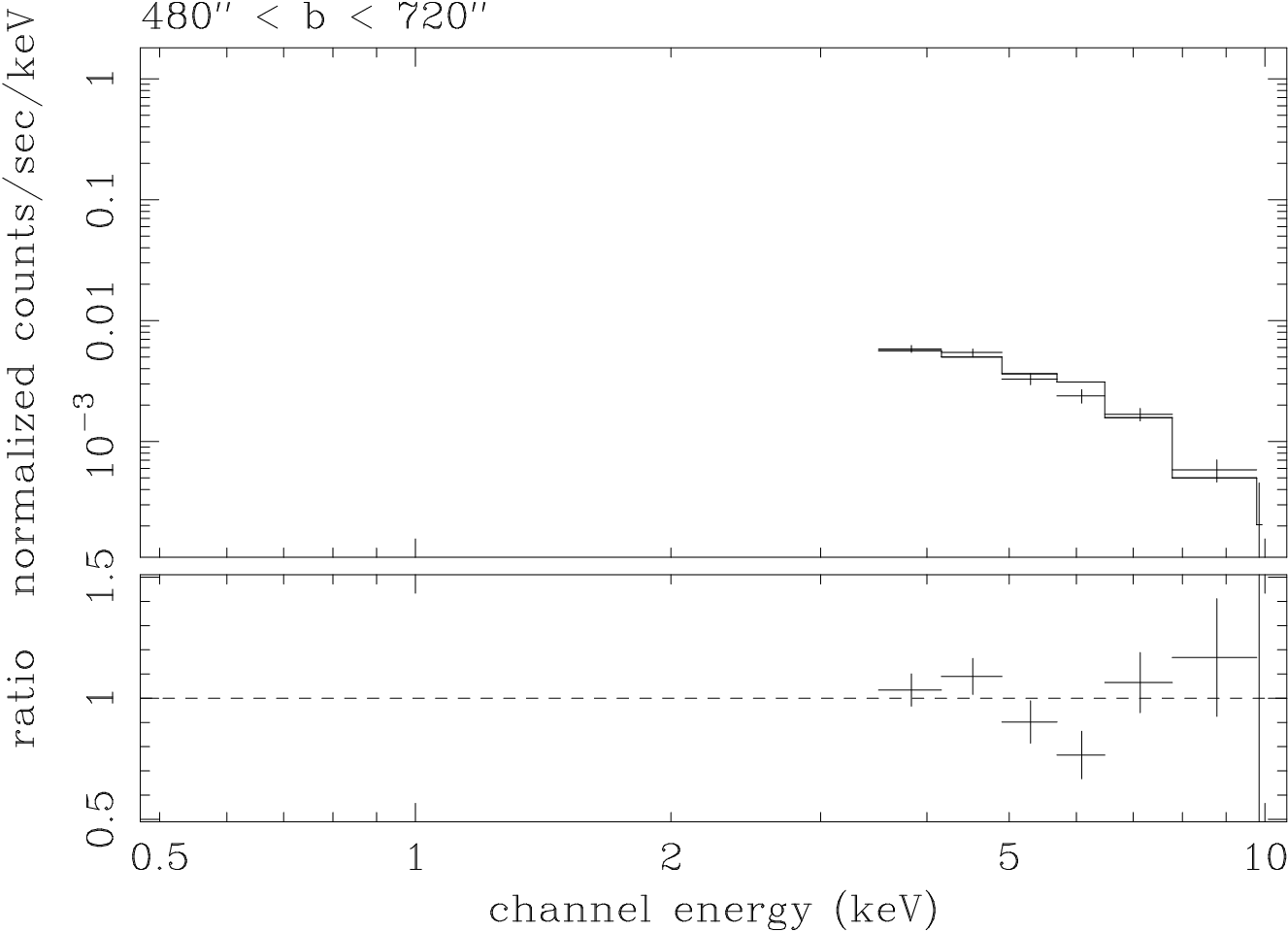}
\includegraphics[angle=-90, width=5cm]{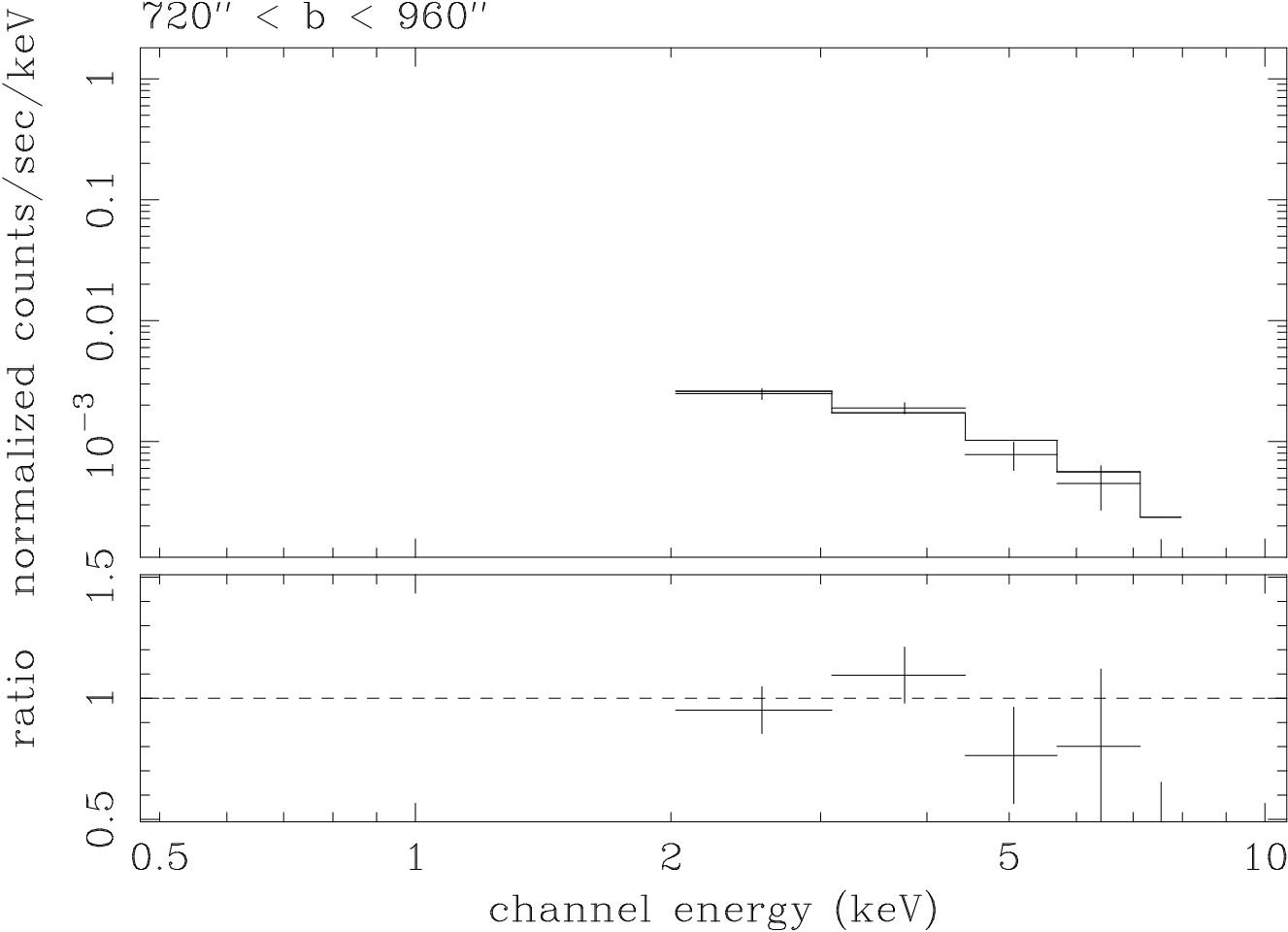}
\figcaption{For any ring, the best-fit spectrum plotted over the data (upper panels). 
The lower panels show the residuals of the model, plotted as ratio data/model. Panels 1-5
refer to {\it Chandra}, 6-9 to {\it XMM-Newton} and 10-13 to {\it Beppo-SAX}. 
\label{fig:ldataratio}}
\end{figure}

The first possibility, discussed in this Subsection, is essentially  a  qualitative examination 
of the residuals left over by the model on the spectrum  of each annulus.
Figure \ref{fig:ldataratio} plots the spectra of the 13 rings together with the
related best fitting models and their  residuals, expressed as ratio data/model.  As expected,
the residuals of {\it XMM-Newton}'s  data are smaller than the others: because of its
wide collecting area, {\it XMM-Newton} strongly dominates the statistics 
of the whole data set.

The qualitative analysis presented here is easily understood recalling how the physical
parameters  shape  the X-ray  spectrum  of a thermal plasma.
The (squared) hydrogen density  provides the
overall normalization of the spectrum, the plasma temperature fixes its  exponential cut
at high energies, and finally the metal abundance gives the  equivalent width of a line.

An inaccurate modelization of  the $n_H$  profile  would appear  as a bad overall normalization 
of the  spectrum of one or more annuli. The lower panels in Figure \ref{fig:ldataratio} show such 
an effect in the innermost  spectrum,  indicating that  our model slightly underestimates the 
central density. The evidence of systematic normalization effects on the outer {\it Beppo-SAX} 
annuli is less clear. 

The systematic residuals at high energy are fairly small for any data set, showing that the 3-D
temperature profile is modeled adequately by our functional form (\ref{eq:kt}).

At the typical temperatures we have determined, the metal contribution mainly 
stems from iron. The local residuals about the 7~keV Fe-K line and the $\simeq 1$~keV Fe-L blend
test  the accuracy  of the determination of the metal  distribution.
The Fe-K line has a good statistics only where the cluster is
hot enough, i.e. in correspondence of the area covered by {\it XMM-Newton}. 
Near the core, where the cluster is cooler, the line emission is dominated by the Fe-L blend.
Unfortunately, the  Fe-L blend  statistics is not high, and {\it Chandra}'s determinations of the metal 
abundance are rather scattered, as shown in Figure  \ref{fig:ab}. In the absence of strong 
constraints from the data, therefore, we conclude that even for the metal distribution  
our model is fairly reliable.

\subsection{Comparing the analytical and geometrical routes: a discussion} 
\label{sec:a1795:comp}

A second way  to check the goodness of our fit is to compare its results with those from 
the  matrix geometrical deprojection. 

\begin{figure}
\epsscale{0.85}
\plotone{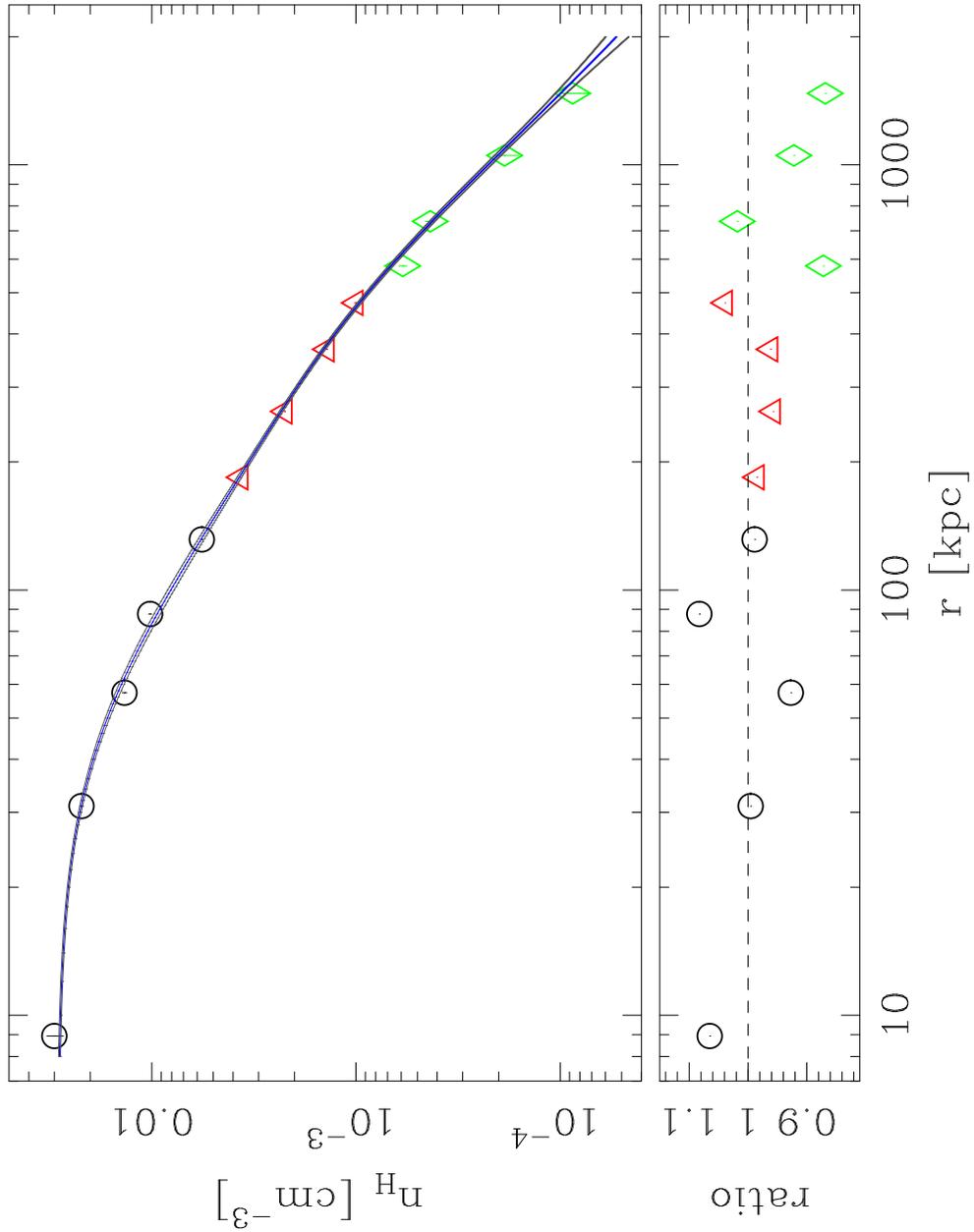}
\figcaption{Upper panel: the deprojected hydrogen density  profile. The central 
continuous curve is the best-fit profile yielded by our analytical 
deprojection with the  double beta-function (see Equation~[\ref{eq:nh}]).
The other curves are the upper and lower boundaries corresponding to
a one-sigma error in the parameters (see text). 
The markers show the data and error bars from  Ettori's matrix geometrical deprojection. 
Circles correspond to {\it Chandra} data, triangles to {\it XMM-Newton}  
and diamonds to {\it Beppo-SAX}.  Lower panel : the ratio between the hydrogen density derived 
from the geometrical and the  analytical deprojection.
\label{fig:nh}}
\end{figure}

\begin{figure}
\epsscale{0.85}
 \plotone{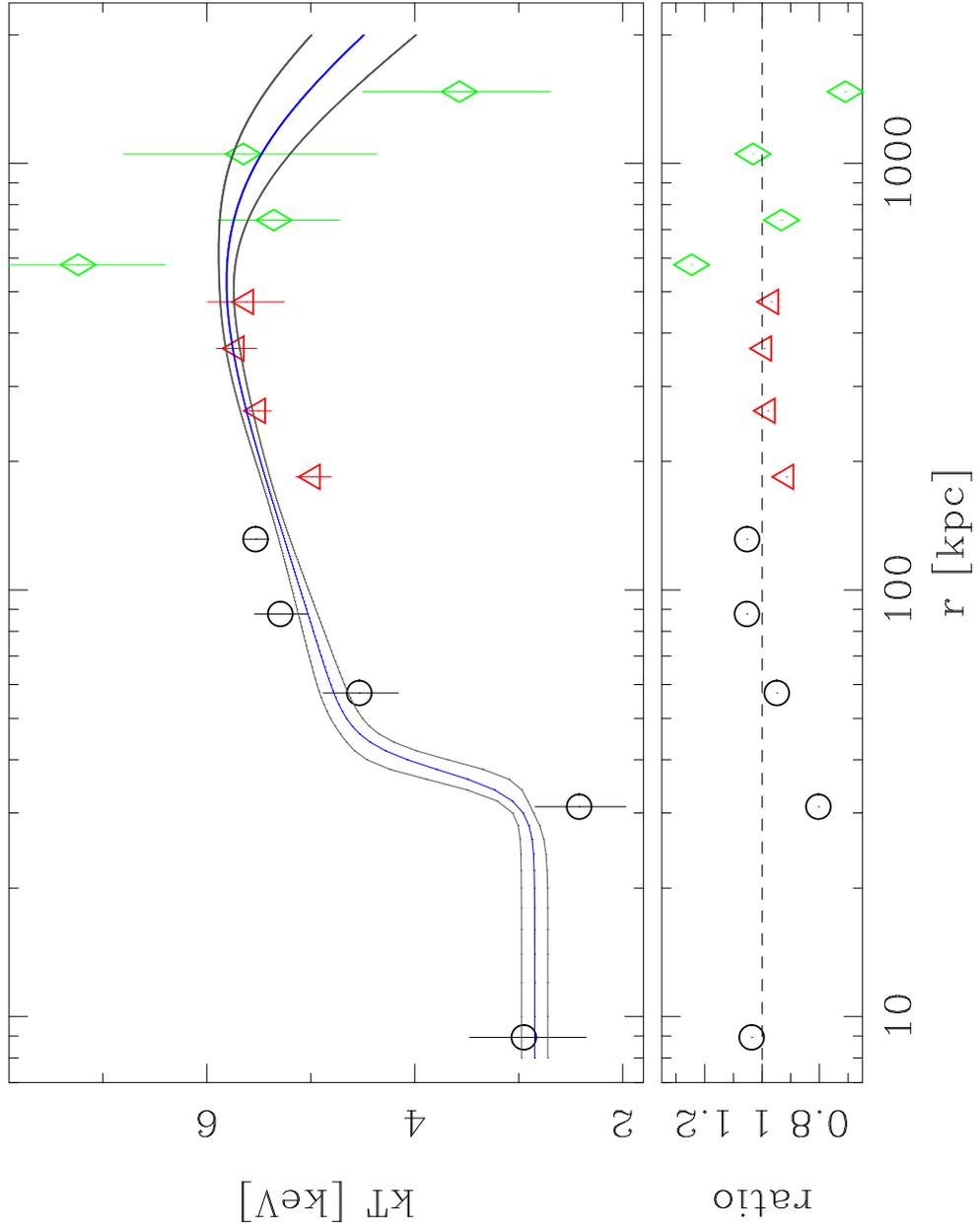}
\figcaption{Upper panel: the  deprojected temperature profile, whose functional form
is given by Equation~(\ref{eq:kt}). The meaning of the symbols is the same as in Figure~\ref{fig:nh}. 
Lower panel : the ratio between the temperature derived from the geometrical and the  analytical 
deprojection.
\label{fig:kt}}
\end{figure}

\begin{figure}
\epsscale{0.85}
\plotone{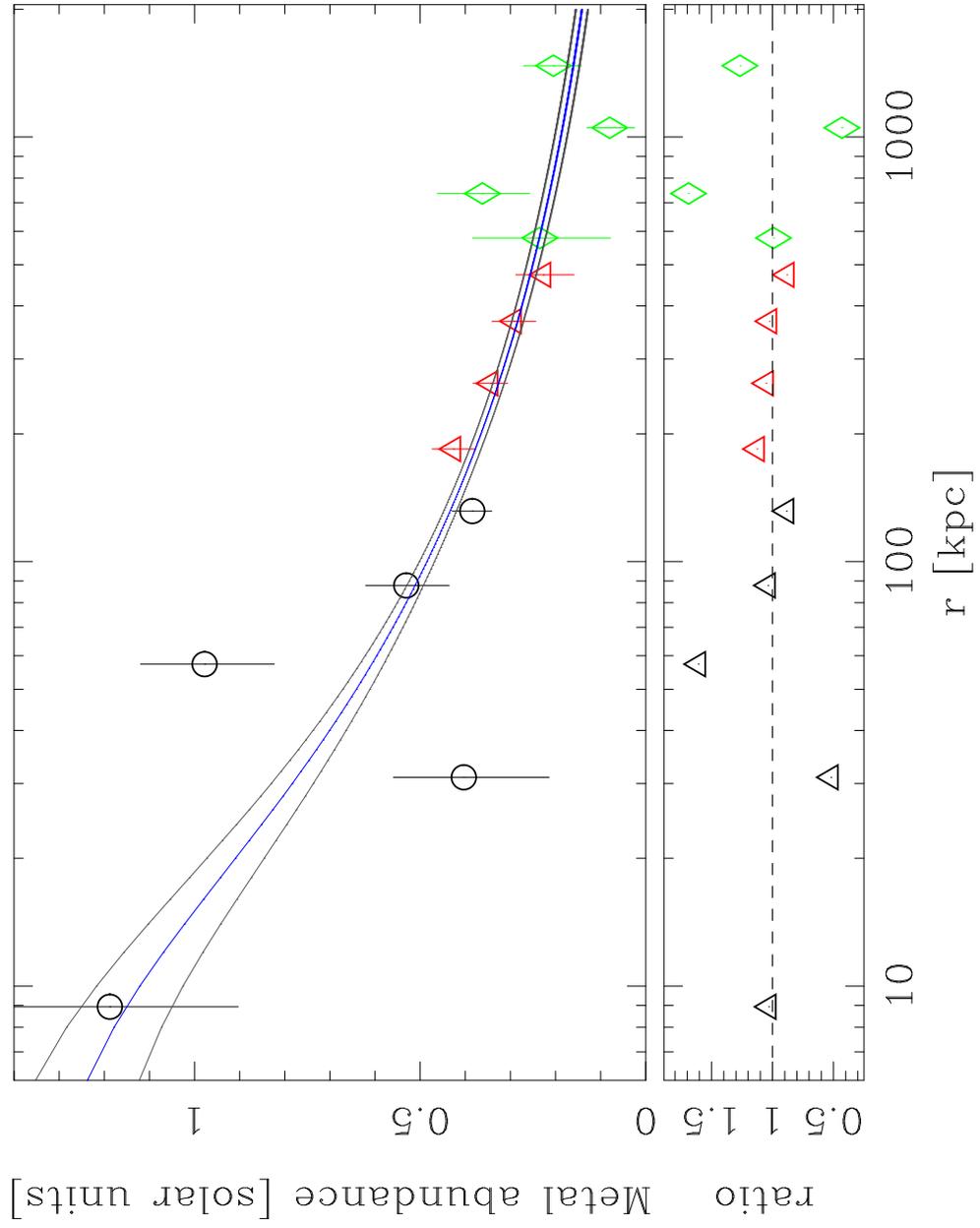}
\figcaption{Upper panel: the deprojected  metal abundance profile, with functional form 
(\ref{eq:ab}). The meaning of the  symbols is the same as in Figures~\ref{fig:nh} and \ref{fig:kt}. 
Lower panel : the ratio between the metal abundance derived from the geometrical and the  
analytical deprojection.
\label{fig:ab}}
\end{figure}

The best-fit analytically deprojected 3-D  profiles of hydrogen density, temperature 
and metal  abundance are plotted in Figures \ref{fig:nh}, \ref{fig:kt},  \ref{fig:ab}, together with 
their geometrically  derived counterparts. Each  of these Figures  also shows the
$1$-$\sigma$ error band about the best-fit profile. The band  enclosing the best-fit curve is the 
area spanned  by all the  functions whose parameters yield a difference  $\Delta \chi^2 \le 1$ 
with respect to the best-fit $\chi^2$. This method of characterizing the errors is  thoroughly 
described in  Appendix C.

Figure  \ref{fig:nh} shows  that our modeled density is smaller than the geometrically deprojected 
value in the central annulus, and larger in the two outermost rings.
Elsewhere the agreement between the two
techniques is good. These results  agree with the qualitative analysis of the
model spectral residuals outlined in the last Subsection. Leaving  aside the outermost 
{\it Beppo-SAX} annuli,  whose
statistics may be weak, the combined results of analytical and geometrical deprojection
seem to suggest that --at least for A1795-- the  density profile is somewhat more peaked than a 
double beta-model. 
 
\begin{figure}
\epsscale{0.85}
\plotone{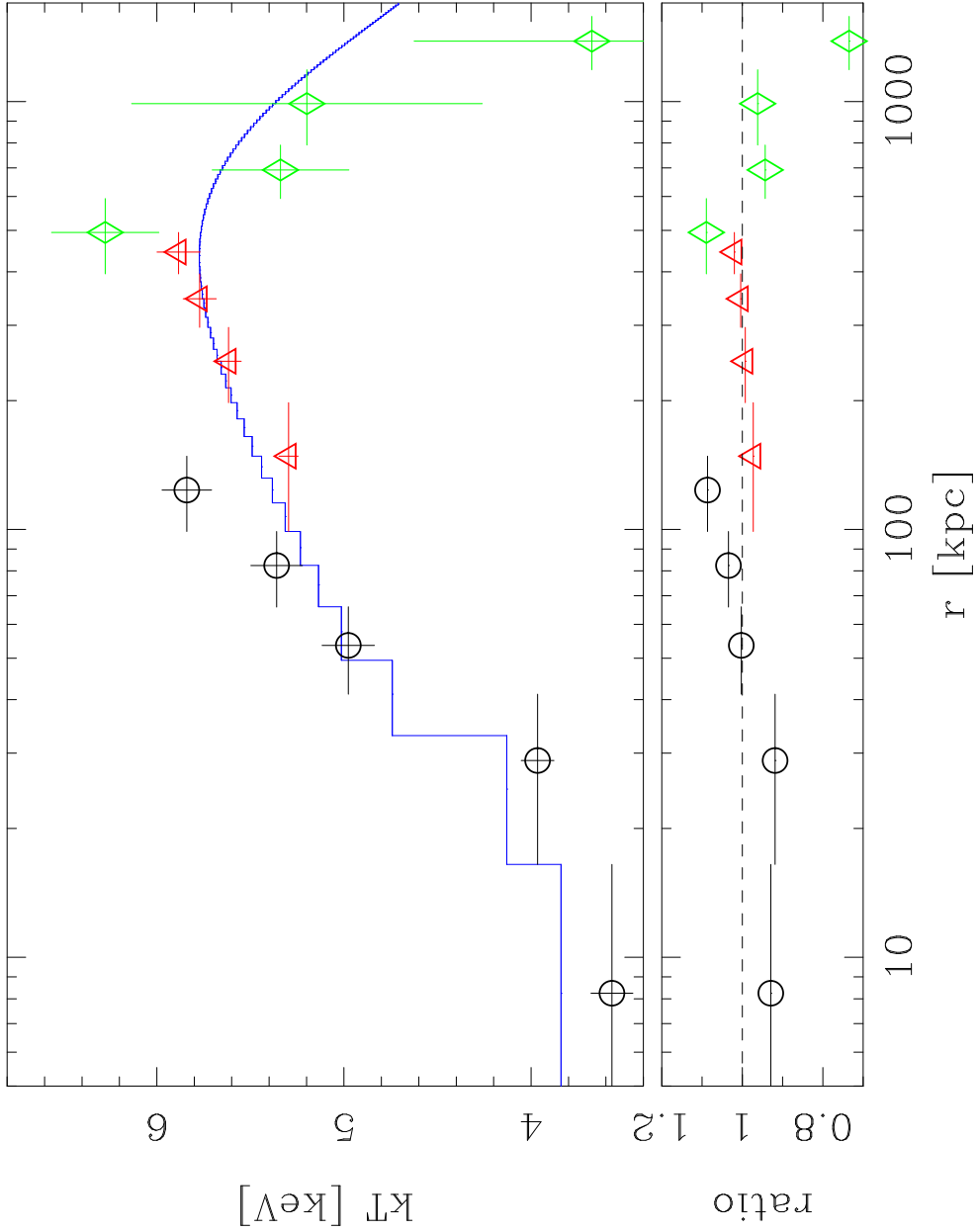}
\figcaption{Upper panel: the re-projected temperature  profile plotted over 
the corresponding  2-D measures provided by the {\sl mekal} analysis.
Lower panel: the ratio between the measured emission-weighted 
temperature and the corresponding re-projected values.
\label{fig:t_2d_3d}}
\end{figure}

It is apparent from Figure \ref{fig:kt} that the 3-D temperature  suffers a sharp discontinuity at  
about $37$~kpc. This fact was anticipated by the strange behavior of the parameter $\varkappa$
noticed  in Section \ref{sec:a1795:anal}. It  is also evident  from the geometrical deprojection, and 
was already noticed by Ettori et al. (2002) in the same position. 
We view the ability of our model to  spot  such an unexpected distribution 
in the three-dimensional temperature profile positively.
Our deprojection finds a maximum 3-D temperature of  $5.8$~keV, 
to be compared  with $7.2$~keV of the geometrical deprojection.  
This difference is due to discrepancies in the  calibration
of the satellites. Figure \ref{fig:t_2d_3d} shows that the temperatures
measured  by  {\it XMM-Newton}  seem too low to match 
well the neighboring values provided by  {\it Chandra}  and {\it Beppo-SAX}. 
Because of their strong statistical weight, {\it XMM-Newton}'s measures 
tend to lower our deprojected profile, reducing also the maximum temperature.

Figure \ref{fig:ab} compares the 3-D metal profiles.
The agreement between the analytical and the geometrical deprojection is good.
In this case it is particularly clear how our profile turns out to be smoother
than its  wiggly  geometrically deprojected counterpart, which would need a regularization
in view of  deriving other physical quantities.

\subsection{Back to Flatland}
\label{sec:a1795:flat}

The third  reliability test we suggest is very stringent, since it 
compares directly the model predictions with the emission weighted, observed quantities.
It consists in  re-projecting  the  3-D best-fit model profiles and comparing  them to the 
2-D measurements. 

The procedure for temperature $T$ and metal abundance $Z$ are the same, 
so we  sketch it  for  temperature only. The temperature $T_{2D}(b_1, b_2)$ averaged over the 
volume seen under the  annulus bounded by the rims $b_1$ and $b_2$ is
\begin{equation*}
T_{2D}(b_1, b_2) = 
\frac{\int_{b_1}^{b_2} db \int_0^\infty dl\, j(r)\, T(r)} {\int_{b_1}^{b_2} db\int_0^\infty dl\, j(r)},
\end{equation*}
where $j(r)$ is the plasma emissivity. The integrals in $dl$ along the
line of sight can be expressed as radial integrals with the geometrical
relation $r^2 = l^2 + b^2$ The resulting double integrals (in $db$ and $dr$) 
can be reduced to single integrals, as described in Appendix A. The previous 
formula is therefore
\be
T_{2D}(b_1, b_2) = \frac{\int_b^\infty dr\, {\cal K}(r, b_1, b_2) \, j(r)\, T(r)} {\int_b^\infty dr\, {\cal K}(r, b_1, b_2) \,  j(r)},
\ee
where the kernel $\cal K$ is given by Equation~(\ref{eq:kern}). As long as the temperature of the ICM 
remains above $\simeq 2$~keV,  the contribution from line emission
is negligible and  the emissivity is mainly due to thermal Bremsstrahlung,
yielding $j \propto n_H^2 T^{1/2}$. We calculate the emission weighted
quantities under this assumption.

\begin{figure}
\epsscale{0.85}
\plotone{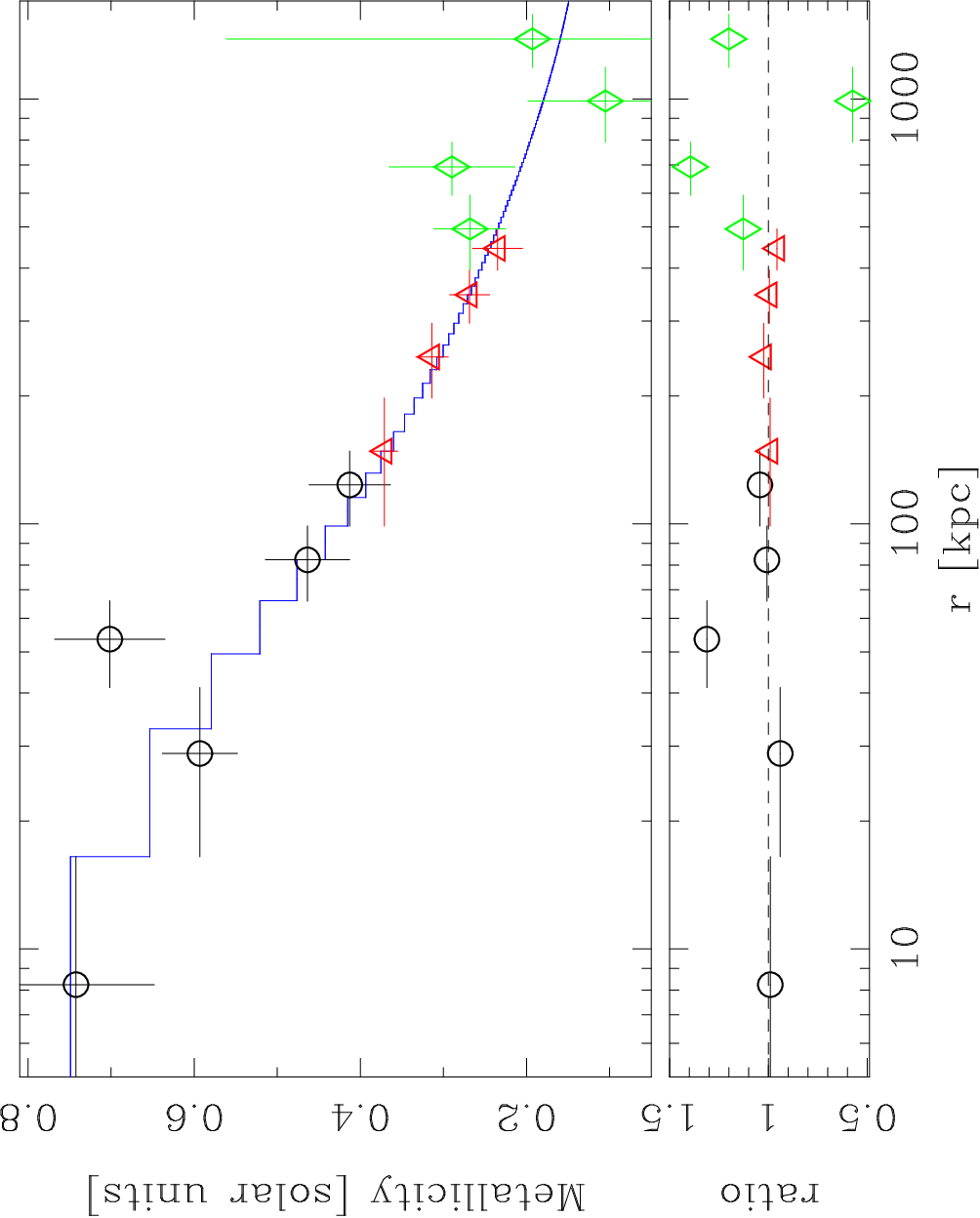}
\figcaption{Upper panel: the re-projected  metallicity profile plotted over
the corresponding  2-D measures. Lower panel: the ratio between the measured 
emission-weighted  metal abundance  and the corresponding re-projected values.
\label{fig:z_2d_3d}}
\end{figure}

Figure  \ref{fig:t_2d_3d} plots  the calculated $T_{2D}(b_1,b_2)$ 
binned over rings  $17$~kpc wide over the  observed values,
directly obtained from the {\sl mekal} analysis.
Similarly,  Figure  \ref{fig:z_2d_3d} compares the
measured and re-projected metal abundance profiles. In both  cases, 
the results are in good agreement with the observations.

\begin{figure}
\epsscale{0.85}
\plotone{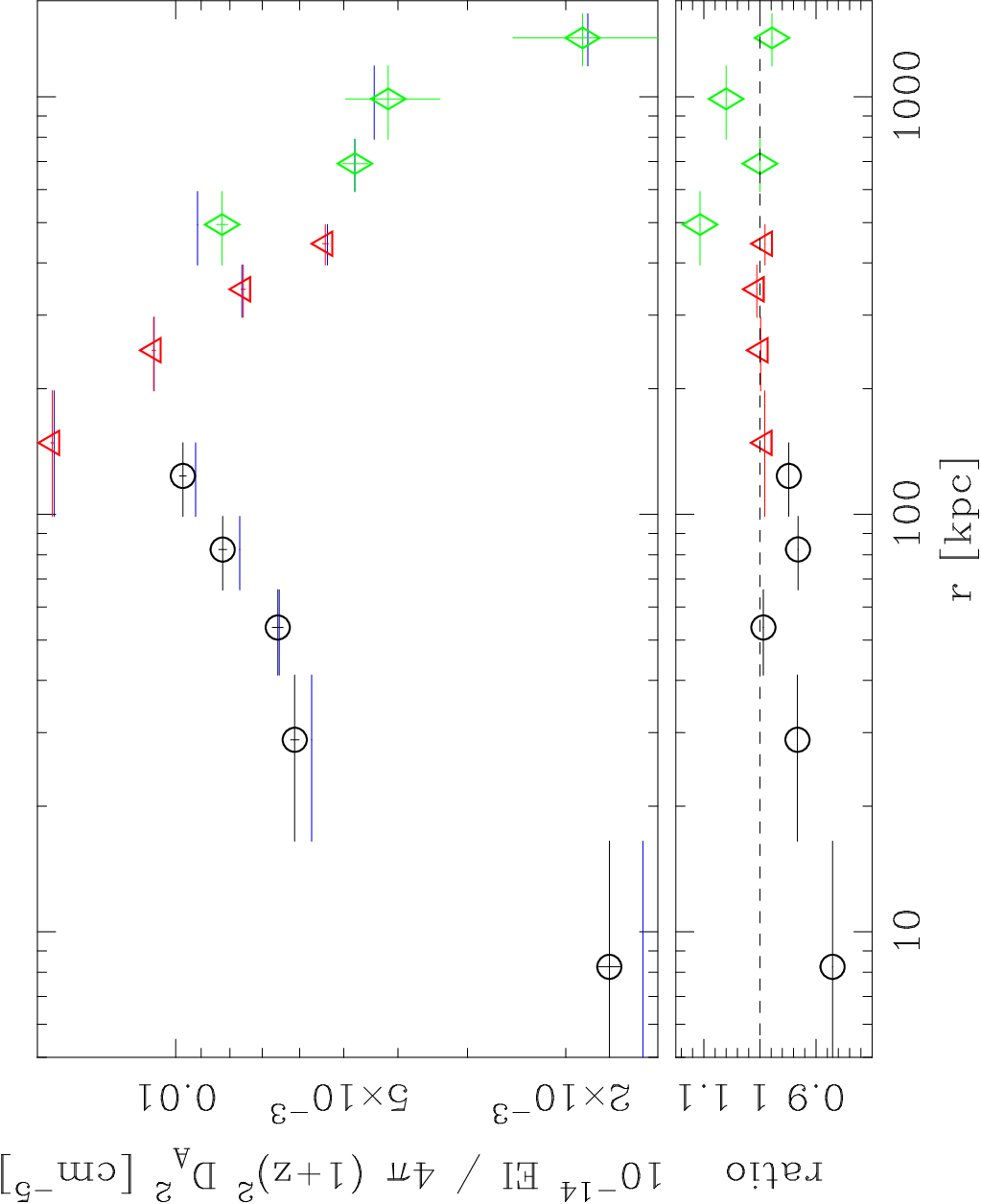}
\figcaption{Upper panel: The re-projected profile of the emission integral 
$EI=\int dV n_e n_H$  plotted over its corresponding  2-D measured counterpart.  
In the {\it y} axis $D_A = 3.4\times 10^2$~Mpc and $z=0.0631$ are 
respectively the  angular distance and the redshift of A1795. 
As before, circles indicates {\it Chandra} data, triangles 
{\it XMM-Newton}  data  and diamonds {\it Beppo-SAX} data. 
The observed quantities stem directly from the normalization of the 
spectra yielded by the {\sl mekal} analysis. The re-projected 
quantities have been calculated by evaluating numerically the emission
integral from the best-fit hydrogen density profile.
Lower panel: The ratio between the {\sl mekal}  emission integral and our reprojected value.
\label{fig:ei_2d_3d}}
\end{figure}

For the re-projection of hydrogen density   we have taken a slightly different route. 
The  geometrical deprojection spectral analysis yields  the {\sl mekal} normalization of the  
spectrum of each ring (Arnaud \& Dorman 2000) 
\begin{equation*}
{\cal M} = \frac{10^{-14} \int dV n_e n_H }{4 \pi D_A^2 (1+z)^2},
\end{equation*}
where the integral is extended over the  volume seen through the ring,
$D_A$ and $z$ are respectively  the angular distance and the redshift 
of the cluster. From our analytical deprojection the distributions  $n_H$ and 
$n_e \simeq 1.21 n_H$  of hydrogen and electron densities are known, and
therefore the quantity ${\cal M}$ can be evaluated for each ring, and compared
to the corresponding value provided by the  {\sl mekal} analysis.
Figure \ref{fig:ei_2d_3d} shows the results, which again are in good  agreement, 
with the possible exception of the innermost ring.

\section{ON THE REJECTION OF AN UNSATISFACTORY  MODEL}
\label{sec:reject}

So far we have devoted a good deal of attention to check the goodness of a single {\it fit}.
In this Section we explore the related issue of how to check the goodness of 
a {\em fitting function}. The problem is to assess our ability to reject an unsatisfactory spatial
profile.  We would like to consider only self-contained tools, which do not 
require to compare  our results with those from other kinds  of deprojection. 

It is clear that the spectral test and the reprojection test discussed above are  well-suited 
for this goal.  A spatial functional form  ought  to be rejected  when it leaves relevant systematic  
errors on many annular  spectra and on the re-projected quantities, and when these residuals 
cannot be reduced with any choice of the functions parameters.
For instance, the reprojection test rules out  an isothermal profile for A1795:  
the  projection  of a constant function  is again a  constant function, clearly at odds with 
the 2-D profile shown in Figure~\ref{fig:t_2d_3d}. The same result comes from the spectral
test, since the systematic residuals left over by an isothermal profile are large.

Both the reprojection and the spectral test have been devised to complement the $\chi^2$ test when
systematic cross-calibration effects due to the use of inhomogeneous data sets are important.
Obviously,  these effects lessen  the reliability of the $\chi^2$ test  also  in 
testing the quality of a fitting function.
There is an important limit, however, in which  the $\chi^2 $ test is still extremely useful, 
namely  when 
the function  is {\it grossly} wrong. In this case, the statistical errors yield a much 
larger  contribution than  the systematics,  and  the $\chi^2$ test is reliable.
For instance, had we taken a simple power-law for the $n_H$ distribution, we would have
obtained  $\chi^2 / {\rm dof} = 62271/3035 \simeq  20.5$.  The   badness of the fit is so marked that 
the value of the $\chi^2$ is only marginally affected by systematic errors, and can be confidently 
used to rule out such an unlikely profile. 

The $\chi^2$ test also rules out both an isothermal, a beta-model  and a  polytropic profile
for the  temperature
\footnote{
 The beta-model for the temperature has been obtained by fixing some parameters in the general 
profile~(\ref{eq:kt}), and corresponds to the function $T=T_0 k(r)$ in the notation of that Equation.
The polytropic profile $T\propto n_H^\alpha$ has been obtained with a slight modification of
the {\sl smaug} code.
}.
The reduced chi-squared is  $\chi^2 / {\rm dof} = 4052 / 3034\simeq 1.3$ for the 
isothermal model, $\chi^2 / {\rm dof} = 3740.6 / 3031\simeq 1.2$ for the  beta-model 
and  $\chi^2 / {\rm dof} =  3781.1 / 3034 \simeq 1.24$ for the  polytropic model.
These figures are not exceedingly larger than our best-fit value, nevertheless the
models  can be safely  rejected with an  F-test, summarized in  Table (\ref{tab:temp}).
As a by-product, these tests indicate that models with few parameters cannot 
reproduce adequately the temperature profile of A1795, and  all the free parameters employed 
to describe our best-fit temperature profile are necessary. 

We have also tested the modified double beta-model
\begin{equation}
n_H  = n_0 \left\{f  [1 + (r/r_{\rm H})^2]^{-\beta_{\rm H}} + 
(1-f)   [1 -  (r/r_{\rm C})^2]^{\beta_{\rm C}} \right\}.
\label{eq:m2b}
\end{equation}
put forward by Ettori (2000) to correct  the original double beta-model, which lacks a satisfactory
physical meaning. 

The rationale of this model is the confinement of the cool gas within the sphere of radius 
$r_{\rm C}$  where  radiative cooling is  important. We have tried several combinations of parameters:
letting all them free, linking or freezing some of them, but in any case the  corrected  double 
beta-model has somewhat worsened the  quality of the fit with respect to the native  double 
beta-model. For instance, by freeing all the density, temperature
and metal abundance parameters (with the exception of the metallicity radius $r_Z$ and the 
temperature exponent $\varkappa$, which is irrelevant to the present discussion), 
the  $\chi^2$ has worsened   from  
$\chi^2 / {\rm dof} = 3652.3 / 3031 \simeq 1.20$ (original double-beta) , to  
$\chi^2 /{\rm dof} = 3785.1 / 3030 \simeq 1.25$ (corrected double-beta).
The best-fit confinement  radius  
$r_{\rm C} = 180$~kpc  is essentially inside  the {\it XMM-Newton} and {\em Beppo-SAX} datasets,
which are therefore described by the hot-beta component alone (the first right-hand term in 
Equation [\ref{eq:m2b}]). In other words, we have given up the degrees of freedom of the cold
component to model the most statistically relevant datasets, which of course reflects on the
final $\chi^2$. 

The use of the $\chi^2$ test to  reject  a model gets difficult  as  the model improves, 
i.e. the functional form  is 
flexible enough to be able to reproduce fairly well the qualitative behavior of the actual 3-D
distributions.  In this case the role of statistical errors decreases in front of the 
systematics, which tend to ``saturate''  the   $\chi^2$,  especially when dealing with inhomogeneous
data sets. In order to get rid of this effect and focus on testing the  reliability of our functional forms,
we have  analyzed  separately  {\it Chandra} and {\it XMM-Newton} datasets. 
In both cases the reduced $\chi^2$ diminishes, ($\chi^2 / {\rm dof}  = 1362.5/1098 \simeq 1.24$ 
for the 5  {\it Chandra} datasets and  $\chi^2 / {\rm dof}  = 1806.2/1535 \simeq 1.18$ for the 4 
{\it XMM-Newton} rings),  but the associated probabilities remains small.   
Since calibration effects are reduced, they are unlikely to be fully responsible for this. 
This indicates   that the introduction of a functional form is
certainly an oversimplification of the complexity of a system like a galaxy cluster. It smoothes out
local inhomogeneities, deviations from spherical symmetry, and so on. It is clear that if the spectra
have  good statistics as in the present case, these unmodeled effects reflect 
on the final  value of  $\chi^2$.

We conclude that a bad  $\chi^2$ model is not always to be discarded. If it has good reprojection
and spectral residuals tests, it should be regarded as a reasonable model for the 3-D structure of
a cluster, and  may be as useful as a
profile obtained via geometrical deprojection. Indeed,  the deprojected profiles are usually taken
as a starting point for other calculations, like the gravitating mass of the cluster.
In a geometrical deprojection one may obtain a good $\chi^2$ statistics on the spectral fits,
but  the  3-D raw profile often needs smoothing to be used for further analyses.
It is clear that smoothing erases  {\it a posteriori} some of the original information.  
Along the analytical route,  some of this information is disregarded {\it a priori}, at the cost of
a relatively high $\chi^2$. The 3-D profiles, however, are smooth and ready for
further uses. Moreover, the knowledge of the functional forms of the distribution in some
occasion may prove useful. 

To summarize,  some warnings are in order: the analytical deprojection is not foolproof: 
a preliminary analysis of the  regions to deproject is necessary to be sure 
not to miss important details of the structure of the cluster to be analyzed. Moreover,
the final fit should never be taken as gospel truth, but always checked against possible flaws
in the model or in the  functional forms for the spatial distributions of the  physical quantities.
To this end, the $\chi^2$ test of the results should be used with care, considering the possible 
role of systematic cross-calibration effects. 
Generally speaking, our advice is to  privilege the spectral and  the reprojection tests  to check 
the quality of  a fit, as well as of the functional form of a  fitting function.

In conclusion, if the analytical method is used carefully, it may provide a reliable deprojection
of a galaxy cluster. Further, the knowledge of explicit functional forms 
for the distributions of density, temperature and metals somewhat simplifies the  after-deprojection 
task of working out  indirect quantities  like the mass  profiles of the cluster,  as 
explained  in the next Section.

\section{RECOVERING THE MASS PROFILES}
\label{sec:gmass}

In this Section we show how the results obtained from analytical deprojection can be used to deduce 
the mass profiles of the cluster.  This part of the procedure is not implemented directly in the 
model {\sl  smaug} within {\it XSPEC},  but it represents   an important complement to the 
information provided by the spectral deprojection. Indeed, it is a sort of {\it a posteriori}
justification of the whole machinery of analytical deprojection.  
Therefore, the codes necessary to this part of the 
spatial analysis of a cluster will be made available to the community to complete {\sl smaug}. 

\begin{figure}
\epsscale{0.85}
\plotone{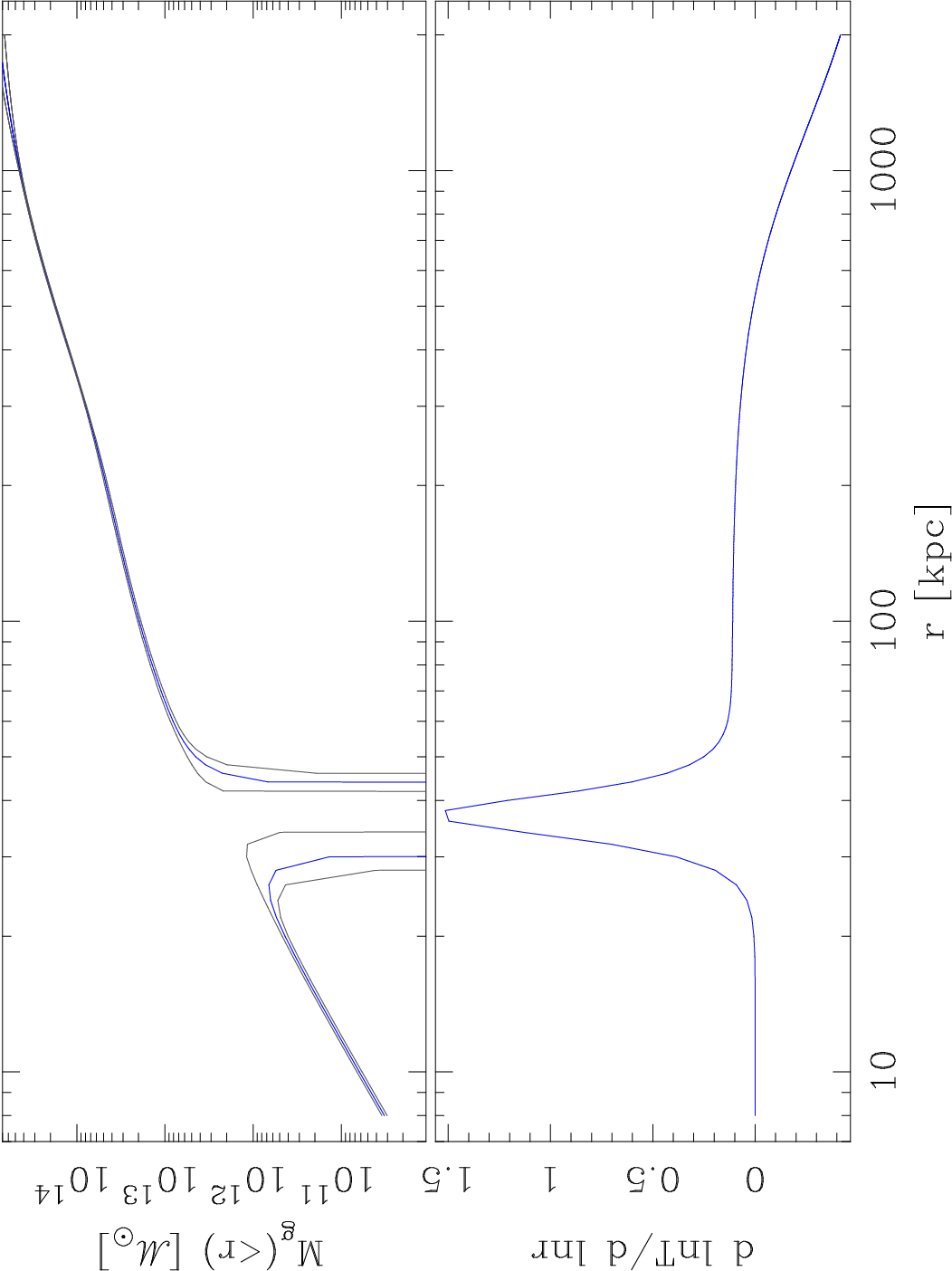}
\figcaption{The gravitating mass calculated with Equation~(\ref{eq:hysta}) (upper panel). 
Because of the discontinuity edge in the temperature profile, the hypothesis of hydrostatic 
equilibrium  breaks down at $\simeq  37$~kpc. This is best appreciated in the lower panel where
we report the logarithmic derivative of the temperature profile.
\label{fig:mglt}}
\end{figure}

\begin{figure}
\epsscale{0.85}
\plotone{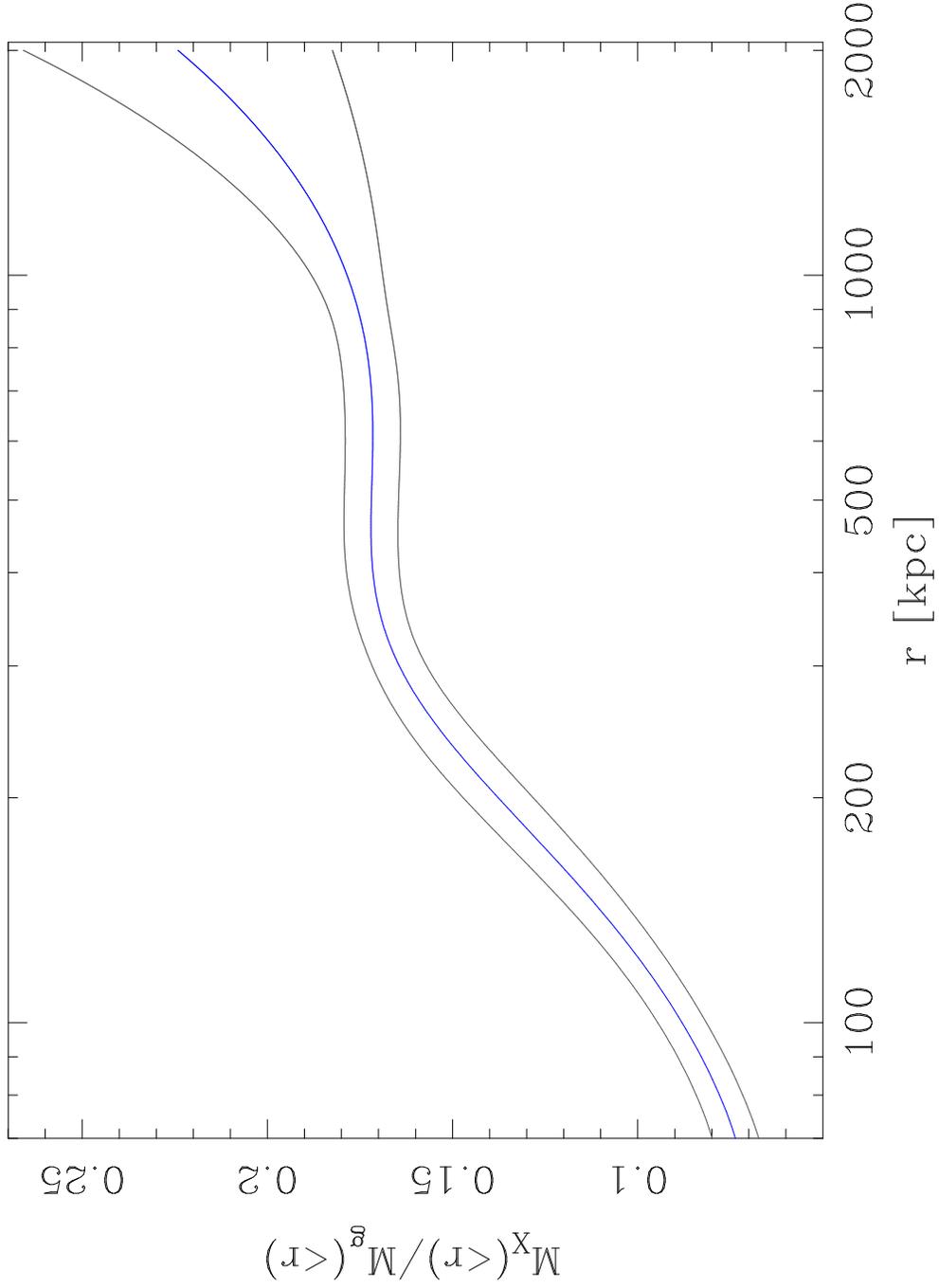}
\figcaption{The X-ray emitting gas fraction within radius $r$ is the ratio of 
the masses given by (\ref{eq:mbar}) and (\ref{eq:hysta}).The plot
excludes the innermost region, where the hydrostatic equilibrium 
fails. The slight rise in the outskirts is statistically
significant only at the $1$-$\sigma$ level, as discussed in Section \ref{sec:gmass}.
\label{fig:bfr}}
\end{figure}

Once the profiles of hydrogen density, temperature and metal   abundance have been determined, 
in principle it is  easy to recover other  interesting physical quantities like the gravitating mass, the 
X-ray emitting gas mass fraction or the metal mass.  This is an advantage of our
analytical deprojection: the  knowledge {\it a priori} of  the functional 
forms of $n_H$, $T$ or $Z$ allows an easy automatization as well as the
numerical stability  of all the necessary procedures.
The hypothesis of hydrostatic equilibrium and the equation of 
state for the ICM  yield the profile of the gravitating mass 
within radius $r$:
\be
\label{eq:hysta}
M_g(<r) = -\frac{k T(r)}{G \mu m_p}\, r\, \left(\frac{d\, \log n_H }{d\,\log r} + \frac{d\,\log T}{d\, \log r}   \right)
\ee
(see e.g. Sarazin 1988). The quantities on the right-hand side of Equation~(\ref{eq:hysta}) 
can be calculated analytically from the profiles (\ref{eq:nh}) and  (\ref{eq:kt}): some lengthy 
but straightforward algebra yields
\begin{equation*}
\frac{d\, \log n_H }{d\,\log r} = \frac{-2}{f\,n_c + (1-f)\,n_g} 
\left[\frac{f \beta_c}{1 +(r_c/r)^2}\,n_c + 
\frac{(1-f) \beta_g}{1 +(r_g/r)^2} n_g 
\right]
\end{equation*}
\begin{multline*}
\frac{d\,\log T}{d\, \log r} =
\frac{-2\, \omega}{1 +(r_t/r)^2} +
\frac{T_1 \, l(r)}{T_0 + T_1\, l(r)} \times \\
\left\{
\frac{\varkappa}{\arctan\left[\left(r/r_{\rm iso} \right)^\varkappa \right]} \; 
\frac{(r/r_{\rm iso})^\varkappa}{1 +(r/r_{\rm iso})^{2\varkappa}}
+ \frac{\xi}{ 1 +(r/r_{\rm cool})^\xi}
\right\}.
\end{multline*}
In the case of A1795, as we have anticipated,  the hypothesis of
hydrostatic equilibrium breaks down at about $37$~kpc from the center,  
because of the discontinuity in
the temperature profile. Mathematically, in Equation~(\ref{eq:hysta}) the 
derivative of $T$ is large and positive, thus making $M_g(<r)$  negative.
This is shown in Figure \ref{fig:mglt}  by the  plots of  the derivative
of $T$ and the  calculated $M_g(<r)$ profile.
It is important to remark  that the hydrostatic determination
of the gravitating mass only fails near the temperature discontinuity. 
The physics of this region is complex: according to 
Markevitch, Vikhlinin \& Mazzotta (2001) the cold gas is slowly moving 
within the 
gravitational potential of the bulk of the cluster, probably after
a past subcluster infall.
Elsewhere the hydrostatic condition holds, and this recipe for
the determination the gravitational mass is relatively safe.
We remark that the ability of our model to spot an unexpected behavior 
of the gravitating mass ought to be considered an index of its flexibility, 
rather than a drawback. 
A simple numerical integration of the hydrogen density (\ref{eq:nh})
provides the mass profile of the X-ray emitting gas enclosed within 
the radius $r$:
\be
\label{eq:mbar}
M_X(<r) \simeq 1.39 \: m_p\: \int_0^r 4\pi\, dr'\, r'^2 n_H(r'),
\ee
where the factor $1.39$ is the correction from the hydrogen to the average ion density.
This mass profile is defined by an integral, thus it is always 
well defined,  unlike the gravitating mass, whose determination relies on
the derivatives of potentially bad-behaved functions, as happens  for A1795. 
Figure \ref{fig:bfr} shows the X-ray emitting gas fraction
$f(<r) = M_X(<r)/M_g(<r)$. It reaches a 
plateau of about $0.17$, but slightly rises in the outskirts.
This final trend (which is nevertheless consistent at the $2$-$\sigma$ level
with being a constant) is related to the possible overestimate 
of the hydrogen  density profile at large radii previously
noticed. This leads to overestimate the gas 
mass profile $M_X(<r)$, and hence the fraction $f(<r) = M_X(<r)/M_g(<r)$ .

\begin{figure}
\epsscale{0.85}
\plotone{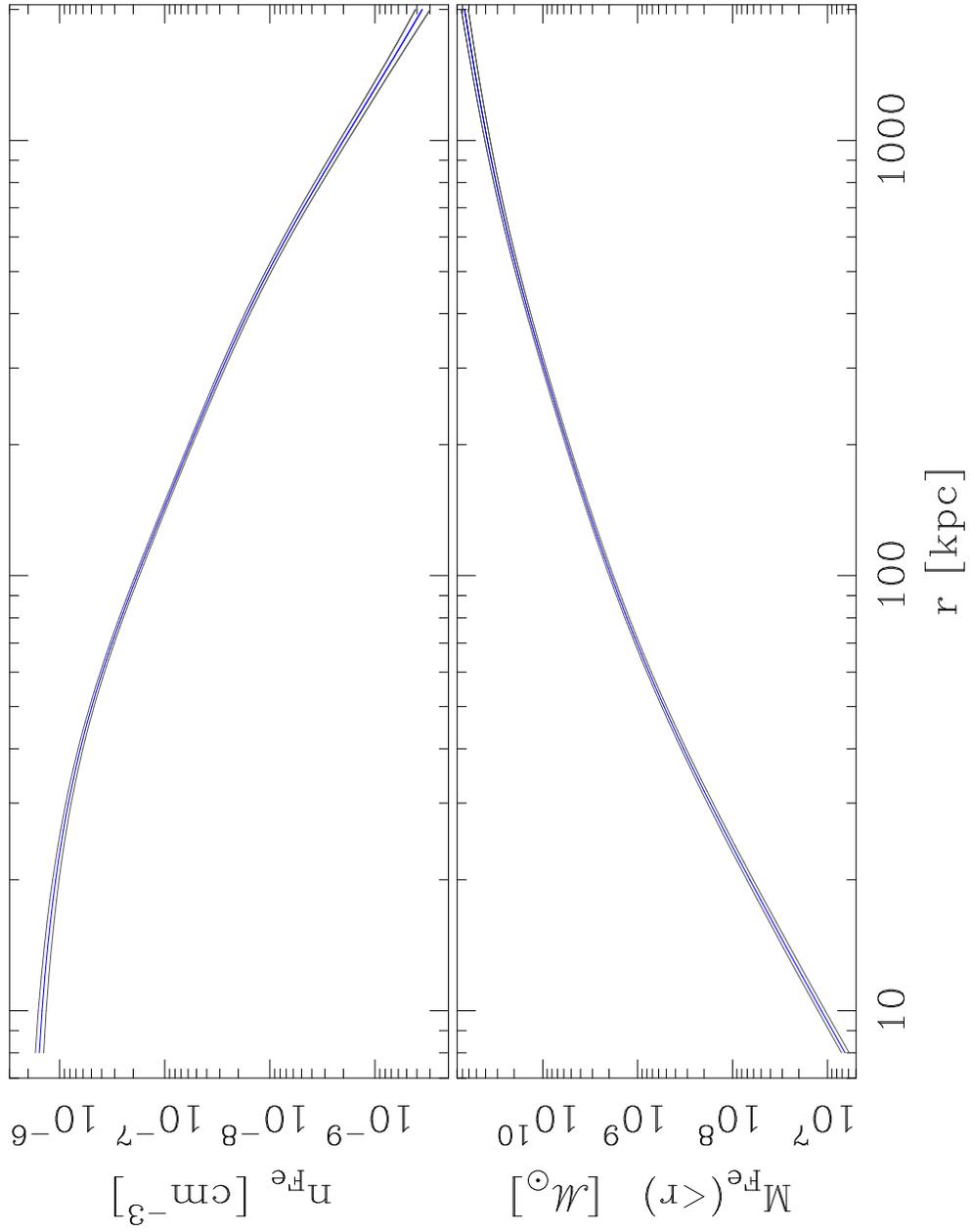}
\figcaption{The iron density (upper panel) and the enclosed iron
mass within radius $r$  (lower panel). The error curves correspond to 
$1$-$\sigma$ uncertainty in the parameters, and have been calculated
as explained in the text (Appendix C).
\label{fig:iron}}
\end{figure}

The combination of the hydrogen and metallicity profiles provides
the metal distribution. At the temperatures observed in  A1795, the main contribution to the 
metallicity profile $Z(r)$ comes from iron, the concentration of which  is approximately
\be
n_{\rm Fe}(r) \simeq  \text{[Fe/H]} \, n_{\rm H}(r) \, Z(r),
\ee
where $[{\rm Fe/H}] = 4.68\times 10^{-5}$ is the solar abundance of
iron with respect to hydrogen, according to  Anders \& Grevesse 
(see Arnaud \& Dorman 2000 and references therein). 
From $n_{\rm Fe}$ it is easy  to deduce 
the enclosed iron mass $M_{\rm Fe}(<r)$. Figure \ref{fig:iron} plots these 
quantities.

\section{SUMMARY}

In this paper we have described a new technique for recovering the
three-dimensional structure of clusters of galaxies. 
The spectrum of a cluster is modeled  as a function of a number of parameters, 
whose values are determined by fitting the model to the observed spectrum. 
This model can be made to work sufficiently fast as a subroutine of  {\it XSPEC} 
because the calculation of the spectral model calls for the evaluation
of a single rather than a double integral (Section \ref{sec:cloud} 
and Appendix A), and because a further analysis has shown that a 
time-consuming  evaluation of the spectral emission function at 
several points throughout the cluster is unnecessary (Appendix B).

The parameters of the model define the known analytical functions of the spatial distributions
of hydrogen density, metal abundance and temperature. From these quantities,  under suitable 
assumptions it is straightforward to obtain other 
physical quantities like the gravitating mass, the X-ray emitting gas  
fraction or  the distribution of metals  within the cluster. 

The distinctive feature  of this technique is the possibility of adopting known analytical fitting 
functions for the use of deprojected data.  After the spectral analysis, no further work is required 
to obtain and  regularize the three-dimensional profiles. Noticeably, our method has a relatively 
simple way to characterize the uncertainty of the results (Appendix C), and does not require a 
Monte Carlo technique. The use of known functions simplifies the reconstruction of
the cluster 3-D structure after the spectral deprojection. Once the best-fit parameters have
been worked out from the spectra --with all the caveats outlined in Section \ref{sec:reject}-- 
this part of the procedure can be easily and safely automatized.

We have shown that the functional forms of  $n_H$, $T$ and $Z$, if suitably 
chosen, may encompass a wide range of possible behaviors of 
the profiles of these physical quantities. This was shown by the ability 
of our model to spot a discontinuity in the temperature profile of A1795.
If the data from the last generation of X-ray satellites required it,
however, it would be extremely simple to modify  these functional
forms to achieve a closer agreement with new observations.

Finally, we have shown that the $\chi^2$ test, the reprojection test and the analysis of the spectral 
residuals left by the models are useful tools to check the reliability of the results, and possibly 
may suggest new and better functional forms for the 3-D distributions of density, temperature  
and metals.

In conclusion, we believe that our deprojection technique
may provide a valid and useful  alternative tool for the deprojection of  galaxy clusters.

\acknowledgements {The authors wish to thank Keith Arnaud and Ben Dorman at 
HEASARC  for their  prompt and kind guide through the mysteries of  {\it XSPEC}. 
Thanks also to our referee Stefano Ettori, whose suggestions greatly
improved the original manuscript.}


\appendix

\section{REDUCTION OF THE DOUBLE INTEGRAL}

In this Appendix we derive Equation~(\ref{eq:spec}) of the text, 
starting from Equation~(\ref{eq:flux}). The integral therein 
is extended over 
the solid angle subtended by an annulus bounded by the rims 
$\bmin$ and $\bmax$. First make the change of variable
from the solid angle to the projected radius
$d\Omega ={2\, \pi b db}/{D_A^2}$, where $D_A$ is the angular distance 
of the source. Then substitute Equation~(\ref{eq:emi}) into (\ref{eq:flux}),
taking into  account the invariance law $I'_{E'}/E'^3 = I_{E}/E^3$ for the 
specific intensity. The result is the photon flux seen by the observer
\be
\label{eq:a1}
{\cal F}_E  = \frac{4\, \pi}{D_A^2\,(1+z)^2}
\int_{\bmin}^{\bmax} db\,b \int_b^\infty dr \frac{r}{\sqrt{r^2-b^2}}\:\epsilon_{E'}(r),
\ee
where the quantity $\epsilon_{E'}=j_{E'}/E'$ is the photon emissivity
at the source, in $\rm photons\,cm^{-3}\,s^{-1}\,keV^{-1}$. 
The primes indicate 
that the energy with respect to the source and the observer's rest frame
differ for the redshift factor $1+z$.
The  integral in (\ref{eq:a1}) can actually be reduced to a single integral.
By inserting a couple of theta functions
$$
\theta(x) =
	\begin{cases}   0 & \text{for $ x < 0$} \\
			1 & \text{for $x  > 0$}
	\end{cases},
$$
the double integral reads
$$
\int_{\bmin}^{\bmax}{ db\,b \int_0^\infty
{ dr \:\epsilon_{E'}(r)\frac{r}{\sqrt{r^2 - b^2}} \theta(r-b)\, \theta(r-\bmin) }}, 
$$
where the order of integration can be reversed; the integral in $db$ can then
be explicitly solved, yielding
$$
\int_{\bmin}^{\bmax}{ db\,b \frac{1}{\sqrt{r^2 - b^2}} \theta(r-b)\, \theta(r-\bmin)} = 
	\begin{cases}
	 (r^2-\bmin^2)^{1/2} & \text{ for  $r <  \bmax$}\\
	(r^2-\bmin^2)^{1/2}\; - \; (r^2-\bmax^2)^{1/2}	 & \text{ for $\bmax<r$}
	\end{cases}
$$
On substituting this in the integral over $dr$, formula
 (\ref{eq:spec}) of the text is  recovered, QED.

\section{NUMERICAL EVALUATION OF THE SPECTRUM}

This Appendix  provides some technical details about the numerical calculation
of the integral in Equation~(\ref{eq:spec}) of the text. 
It is useful to single out the hydrogen density in the emissivity function:
$$
\epsilon_E = \frac{j_E}{E} = n_e n_H \frac{\Lambda_E(T,Z)} {E} \equiv 
n_H^2 \lambda_E(T,Z).
$$
Omitting irrelevant factors, the  integral to be calculated is 
\be
{\cal F}_E = \int_{\bmin}^\infty \, dr{\frac{d \phi}{d r} \lambda_E(T,Z)},
\label{eq:b1}
\ee
where the differential emission  integral 
$$
\frac{d \phi }{d r} = {\cal K}(r; b_{\rm min},b_{\rm max})\;  n_H^2(r),
$$
is far more sensitive to the radius $r$ than the  spectral  function 
$\lambda_E(T,Z)$, which depends implicitly on $r$ through 
the functions  $T(r)$  and $Z(r)$.
This suggests to split the integration interval in several isothermal shells 
with homogeneous metallicity. Therefore, in  each shell the 
function $\lambda_E(T,Z)$ needs only one evaluation. A high numerical 
accuracy is put only in 
evaluating the emission integral of the shell. This approach greatly
reduces  the (computationally heavy) evaluation of $\lambda_E(T,Z)$ 
in several mesh points, but preserves the accuracy of the results.

The range of integration is cut-off with the introduction of a 
user-defined maximum radius  $r_{\rm max}$, typically in the order
of $\simeq 2$~Mpc, and the interval $[b_{\rm min}, \,r_{\rm max}]$ 
is split into  $N$ shells bounded by the nodes $r_0,r_1,\cdots .r_N$ 
(in general not evenly spaced); the extremes $r_0$ and $r_N$ 
coincide with  $b_{\rm min}$ and  $r_{\rm max}$ respectively.
Equation~(\ref{eq:b1}) is rewritten as
\begin{eqnarray}
\left\{
	\begin{array}{l}
	{\cal F}_E = \sum_{i=1}^N{ \phi(i)\; \overline{\lambda_E(i)}}\\
\\	
	 \phi(i) =  \int_{r_{i-1}}^{r_i} { dr\; \frac{d \phi}{d r}}\\
	\end{array}
\right.,
\label{eq:b2}
\end{eqnarray}
where the average spectral  function $\overline{\lambda_E(i)} $ over 
the $i$-th emitting shell is {\it defined} as 
\be
\label{eq:ave}
\overline{\lambda_E(i)} = \frac{\int_{r_{i-1}}^{r_i}\,dr\: \frac{d \phi}{d r} \lambda_E}{\phi(i)}.
\ee
Up to now the equation are {\it exact}: the approximations are to
be introduced now in the evaluation of  the average spectral function
$\overline{\lambda_E}$.
For the sake of readability of the formulae we drop both the spectral 
index $E$ and the label of the volumes.
In Equation~(\ref{eq:ave}) we expand $\lambda(T,Z)$ in a neighborhood of the 
(still undefined) average temperature and metallicity $\tew$ and 
$\zew$, defined on each shell:
$$
\lambda (T,Z) = \lambda(\tew,\zew) + \frac{\partial \lambda}{\partial T}(T-\tew) +  \frac{\partial \lambda}{\partial Z}(Z-\zew) + {\cal O}^2.
$$
The term ${\cal O}^2$ denotes higher-order terms in $(T-\tew)$ and $(Z-\zew)$.
If we put this expansion in the right-hand side of (\ref{eq:ave}) 
we obtain 
\be
\overline{\lambda} = \lambda(\tew,\zew) + \frac{\partial \lambda}{\partial T}
\left[
\frac{\int dr \frac{d\phi}{ dr}\,T} {\phi} - \tew
\right]
\:
+
 \frac{\partial \lambda}{\partial Z}
\left[
\:
\frac{\int dr \frac{d\phi}{ dr}\,Z}{ \phi } - \zew
\right]
\:
+
\:
{\cal O}^2.
\ee
If we  define  the shell averages $\tew$ and $\zew$ as
\begin{eqnarray}
\label{eq:ew1}
\tew \equiv \frac{\int dr \frac{d \phi}{ dr} \: T}{ \phi } \\
\zew \equiv \frac{\int dr \frac{d \phi} {dr} \: Z}{ \phi },
\label{eq:ew2}
\end{eqnarray}
we find $\overline{\lambda} = \lambda(\tew,\zew) + {\cal O}^2 $, 
i.e. the (approximated) spectral function evaluated in  $\tew$ and $\zew$ 
is first-order accurate with respect to the (exact) 
spectral function (\ref{eq:ave}), emission-weighted within a shell.  

The trapezoidal rule is accurate enough for the calculation of the integrals 
in the definitions (\ref{eq:ew1}) and (\ref{eq:ew2}): 
on the shell $[r_i, r_{i-1}]$, for instance 
$$
\tew_i \simeq \frac{{\cal K}(i)\, n_H^2(i)\, T(i)\; +
\; {\cal K}(i)\, n_H^2(i-1)\, T(i-1)}{{\cal K}(i)\, n_H^2(i) \;+\; {\cal K}(i-1)\, n_H^2(i-1)},
$$
where ${\cal K}(i)$, $T(i)$, and $n_H(i)$ denote the values of the functions 
$\cal K$, $T$, and $n_H$ evaluated at $r_i$.
As mentioned before, the evaluation of the  emission integral $\phi(i)$ 
in the sum (\ref{eq:b2})  calls for more accuracy, and therefore we adopt 
a 10-point Gauss-Legendre formula (see e.g. Press et al. 1992).
In  each integration shell the differential emission integral is
calculated  10 times, which is accurate and relatively fast; the evaluation 
of  the spectral function  $\lambda_E(T,Z)$, 
which is substantially slower, occurs only once.

A final remark is in order about the spacing of the integration nodes
$r_0,r_1,\cdots .r_N$. Since the differential emission integral usually
strongly peaks at $r\lesssim b_{\rm max}$, we have chosen 
a quadratically spaced grid, which yields a  better accuracy than an evenly
spaced one.

\section{ON THE CHARACTERIZATION OF THE ERRORS}

In  this Appendix we sketch  our method  to characterize the error on a 
function $f(r;{\mathbf a})$ dependent on the radial coordinate $r$ as well
as  on  the $N$ parameters  $\mathbf{a} = (a_1, \cdots, a_N)$. 
Our  problem is to show how their uncertainties reflect on the 
function $f$. We start by considering the Taylor expansion of the 
$\chi^2$ about the best-fit values of the parameters 
$\mathbf{\tilde{a}} = (\tilde{a}_1, \cdots, \tilde{a}_N)$.
Here $\chi^2$ has a minimum, and its expansion to the lowest order reads
\be
\label{eq:curv}
\Delta\chi^2 \equiv \chi^2({\mathbf a}) - \chi^2({\mathbf{\tilde a}}) =
\frac{1}{2}\sum_{i, k = 1}^{N}
 (a - \tilde a)_i (a-\tilde{a})_k \frac{\partial^2\chi^2(\mathbf{\tilde a})}{\partial a_i \partial a_k}.
\ee 
In the same fashion, 
the first-order Taylor expansion of $f(r;{\mathbf a})$ about 
$\mathbf{\tilde{a}}$ reads
\be
\label{eq:hpla}
\Delta f \equiv f(r;{\mathbf a}) - f(r;\mathbf{\tilde{a}}) = \sum_{i = 1}^{N } (a - \tilde a) _i
\frac{\partial f(r;\mathbf{\tilde{a}})}{\partial a_i}.
\ee
In what follows we assume  that $r$ is fixed, and focus our attention 
on the $N$-dimensional
parameter space. Here the manifold  $\Delta\chi^2$~=~constant is the 
($N-1$)~-~dimensional surface $\cal{H}$ of an ellipsoid. 
In the same space, Equation~(\ref{eq:hpla}) defines a family 
$\cal{P}$ of parallel planes  dependent on the intercept $\Delta f$.
The intersection set $\cal{H} \cap \cal{P}$  contains  the 
parameters which  for the assigned  value of $\Delta\chi^2$
give the  absolute difference $|\Delta f|$ with respect to the best-fit 
determination of $f$. If  $|\Delta f|$ is 
large, the intersection  is empty: the error on $f$ is too large to be 
consistent with the given $\Delta\chi^2$. On the other hand, 
if $|\Delta f|$ is small, 
there are infinitely many points in the intersection, i.e. several 
combinations of parameters are able to define  many functions characterized 
by the same  $\Delta\chi^2$. It should be clear (and may be proved rigorously) 
that the largest 
value of $|\Delta f|$   compatible with the fixed  value of $\Delta\chi^2$ 
occurs for  the values  $\mathbf q$ of the parameters for which the ellipsoid 
 (\ref{eq:curv}) and one plane of the family  (\ref{eq:hpla}) are tangent.
Quite generally, given a point $T$ of coordinates $\mathbf q$ 
lying on the surface $\Delta\chi^2$~=~constant of the ellipsoid 
(\ref{eq:curv}), the plane tangent to the ellipsoid at $T$ is
\be
\label{eq:tan}
\sum_{i, k = 1}^N ({\cal E}^{-1})_{i\,k} (q - \tilde a)_i (a - \tilde a)_k = \Delta\chi^2,
\ee
where the error matrix ${\cal E}_{i\,k}$ is the inverse of the matrix 
defined by
$$ 
({\cal E}^{-1})_{i\, k} =\frac{1}{2} \frac{\partial^2\chi^2(\mathbf{\tilde a})}{\partial a_i \partial a_k}.
$$
The tangent plane (\ref{eq:tan}) at $T$ coincides 
with the plane of the family (\ref{eq:hpla}) passing through
$T$ if the coefficients of both are proportional to each other. This 
condition yields
\be
\label{eq:df}
{\Delta f}^2 = \Delta\chi^2\, \sum_{i, k =1}^N {\cal E}_{i\, k } \frac{\partial f}{\partial a_i}  \frac{\partial f}{\partial a_k}. 
\ee
Since  ${\cal E}$ is the reciprocal of a (symmetrical)  Hessian matrix 
evaluated at a local (and hopefully, absolute) minimum of $\chi^2$, it 
is symmetrical and positive definite;
the equation for $\Delta f$  has two roots $\pm |\Delta f|$.
As the radial coordinate  $r$ varies,  $f(r) \pm \Delta f (r)$  draws the  
error band  associated with  the given value of  $\Delta\chi^2$. 

Alternatively, formula (\ref{eq:df}) can be obtained by maximizing the
value of $|\Delta f|$ (as a function of the parameters $\mathbf a$), 
subjected to the condition that  $\mathbf a$ belongs to the
assigned manifold $\Delta\chi^2$~=~constant. An application of  the standard
Lagrange multipliers technique leads again to Equation~(\ref{eq:df}).

In Equation~(\ref{eq:df})  the components of 
$\nabla_{\mathbf a} f(r;\mathbf{\tilde{a}})$ are known
from the functional form of $f$ and from the best-fit values of the 
parameters.  The matrix  ${\cal E}$ is {\it not} directly known, but it can 
be worked out using the variances and the 
principal axes returned  by {\it XSPEC} at the end of the fit procedure.
The orthogonal matrix $\cal{V}$  of the principal axes is made up by the (row)
 eigenvectors of ${\cal E}$,
and the variances are the corresponding eigenvalues. Therefore, ${\cal E}$ 
is given by 
${\cal E} = {^t}{\cal V} \Sigma {\cal V}$, where  $\Sigma$ is the  matrix with 
the variances on its  main diagonal, and ${^t}{\cal V}$ is the transpose 
matrix of ${\cal V}$ (the interested reader is referred to 
any textbook of linear algebra, e.g. Lang 1966).

The next degree of complication occurs when more than one function is present,
as it is our case (the reader is  reminded that we determine simultaneously 
the functions $n_H$, $T$ and $Z$).
Suppose that besides $f$ there is another function $g$: for instance $f$ and 
$g$  could be the temperature and the hydrogen density. In order to avoid 
unnecessary complications, suppose that each function  depends  only on  
the radius $r$ and on one parameter: $f=f(r,a)$ and $g=g(r,b)$.
The parameter space is the two-dimensional plane $ab$, the error 
matrix has rank 2, the $\chi^2$ depends both on $a$  and on $b$ and the 
surface  $\Delta \chi^2$~=~constant is an ellipse in the plane $ab$.
Assume for the moment
that  $a$ and  $b$ are mutually independent, i.e. they are not linked to each
other in the fitting procedure: notice however that this by no means entails they are statistically 
uncorrelated. As the parameter $a$ varies about its best fit value $\tilde{a}$,
the error $\Delta f$  also varies, and in the same fashion as before its 
largest value 
\be
\label{eq:erf}
{\Delta f}^2 = \Delta\chi^2\,  {\cal E}_{1\, 1 } \left(\frac{\partial f}{\partial a}\right)^2
\ee
occurs where the error ellipse and the line
$$
\Delta f = \left.\frac{\partial f}{\partial a}\right|_{a = \tilde{a}} (a -  \tilde{a})
$$	
are tangent. Similarly, the error on $g$ is
\be
\label{eq:erg}
{\Delta g}^2 = \Delta\chi^2\,  {\cal E}_{2\, 2 } \left(\frac{\partial g}{\partial b}\right)^2.
\ee
It might be objected that this procedure does not take into proper account 
a possible statistical correlation between $a$ and $b$. This is expressed by
the off-diagonal terms of the error matrix, that however do not  appear in 
formulae (\ref{eq:erf}) and (\ref{eq:erg}). 
This objection is incorrect, 
since the correlation between $a$ and  $b$ (if any) intervenes also
through the diagonal terms ${\cal E}_{1\, 1}$ and ${\cal E}_{2\, 2}$. Indeed,
the correlation introduces a tilt in
the principal axes of the error ellipse with respect to the coordinate 
axes of the parameter plane. Therefore the diagonal terms of the error matrix
in Equations~(\ref{eq:erf}) and (\ref{eq:erg}) differ in the cases of
correlation or uncorrelation between the fit parameters $a$ and $b$. We may
conclude that the errors $\Delta f$ and $\Delta g$ are  correctly calculated 
in both cases.

The situation is slightly more complicated if an explicit link 
$b = \phi(a)$ is introduced in the fit between $a$ and $b$; $b$ is no more a free fit
parameter, and its role is taken over by $a$. Because of the link, the quantity
$\partial g/\partial a$ does not  vanish, since
\be
\frac{\partial g}{\partial a} = \frac{\partial g}{\partial b} \phi'(a).
\ee
The formula (\ref{eq:erg}) in this case is replaced by
\be
\label{eq:erg2}
{\Delta g}^2 = \Delta\chi^2\, \left\{ 
{\cal E}_{1\, 1} \left(\frac{\partial g}{\partial b}\, \phi'\right)^2 + 
2\,{\cal E}_{1\, 2}\frac{\partial g}{\partial b}\, \phi' + 
{\cal E}_{2\, 2} \left(\frac{\partial g}{\partial b}\right)^2
\right\}.
\ee
This procedure of dealing with the links between parameters allows to
derive  the error for one special parameter, namely the hydrogen central
density $n_0$ of the model {\sl smaug}. $n_0$ is special because during
the fit it is {\it formally} fixed, but its role is played by the 
normalization  $\cal N$ of  {\sl smaug}, being ${\cal N} = n_0^2$. We 
interpret this relation as a formal link 
$n_0 = \phi({\cal N}) = \sqrt{\cal N}$ between ${\cal N}$ and $n_0$.
In Equation~(\ref{eq:df}), the derivative with respect to the fit parameter 
${\cal N}$  of a function $f$ dependent on $n_0$ reads
\be
\frac{\partial f}{\partial {\cal N}} = \frac{\partial f}{\partial n_0} 
\frac{\partial n_0}{\partial {\cal N}} 
= \frac{1}{2\cal{N}}\, \frac{\partial f}{\partial n_0}.
\ee
As a final point, let us consider how it is possible to characterize the
error of a function $h = h(f(r,a), g(r,b))$ dependent both on $f$ and 
$g$: for instance, if $f$ and $g$ are the temperature and the hydrogen 
density $h = f\,g$ is the pressure. The basic formula is always (\ref{eq:df}),
and therefore it is only necessary to evaluate correctly the derivatives
on its right hand side. If the fit parameters $a$ and $b$ are not linked, it is
$\partial h/\partial a = (\partial h/\partial f) (\partial f/\partial a)$
and $\partial h/\partial b = (\partial h/\partial g)(\partial g/\partial b)$.
The occurrence of a link $b = \phi(a)$ leads to
\be
\frac{\partial h}{\partial a} = \frac{\partial h}{\partial f} \frac{\partial f}{\partial a} +   \frac{\partial h}{\partial g} \frac{\partial g}{\partial b}\phi'(a).
\ee

\newpage


\newpage

%
%
%
%
%
%
%
%
%
%
%
%
%
%
%
%
%
%
%
%
%
%
\newpage

%
%
\begin{deluxetable}{cccccc}
\tabletypesize{\small}
\tablecaption{\sc Extraction regions for the spectra
\label{tab:summ}}
\tablehead{
	\colhead{Instrument}            & 
	\colhead{Cross}                 & 
	\colhead{Spectral range}        &  
	\colhead{Inner ring}            & 
	\colhead{Outer ring}            & 
	\colhead{Shading}        \\
	\colhead{}			&
	\colhead{normalization}		&
	\colhead{[keV]}			&
	\colhead{[arcsec]}		&
	\colhead{[arcsec]}	        &
	\colhead{factor}
}
\startdata
\multirow{5}{30mm} {\it Chandra} 
        &  \multirow{5}{30mm}{106$\%$}   &  0.5--7.0   &    0   &   10  &  1.000  \\
 	&                                &  0.5--7.0   &   10   &   25  &  1.000  \\
	&                                &  1.0--7.0   &   25   &   40  &  1.000  \\
	&                                &  1.0--7.0   &   40   &   60  &  1.000  \\
	&                                &  1.0--7.0   &   60   &   90  &  1.000  \\
\hline
\multirow{4}{30mm} {\it XMM-Newton}
	&  \multirow{4}{30mm}{100$\%$}  &  1.0--10.0   &   60   &  120  &  1.000  \\
	&                               &  1.0--10.0   &  120   &  180  &  1.000  \\
	&                               &  1.0--10.0   &  180   &  240  &  1.000  \\
	&                               &  1.0--10.0   &  240   &  300  &  0.981  \\
\hline
\multirow{4}{30mm} {\it Beppo-SAX}
	&   \multirow{4}{30mm}{95$\%$}  &  2.0--10.0   &  240   &  360  &  1.000  \\
	&                               &  2.0--10.0   &  360   &  480  &  0.981  \\
	&                               &  3.5--10.0   &  480   &  720  &  1.000  \\
	&                               &  2.0--8.0    &  720   &  960  &  0.794  \\
\hline
\enddata
\tablecomments{The cross-normalization constants  among the instruments are free 
parameters of the model. The value relative to {\it XMM-Newton}
is taken as unity by  definition. The shading  factor takes into account 
the reduced area of the extraction
region of two {\it Beppo-SAX} spectra due to the entrance window.}
\end{deluxetable}
%
%
\begin{deluxetable}{clrc}
\tabletypesize{\small}
\tablecaption{\sc Results for the {\sl smaug} model.\label{tab:best}}
\tablehead{
	\colhead{Free parameter}       & 
	\colhead{Best-fit value}       & 
	\colhead{One-sigma error}     &  
	\colhead{Units}              
}
\startdata
$f$            &   $4.41 \times 10^{-2}$  & $1.40 \times 10^{-2}$   &  --                \\ 
$n_0$          &   $2.89 \times 10^{-2}$  & $8.91 \times 10^{-4}$   &   {\rm cm}$^{-3}$  \\ 
$\beta_c$      &   $1.76 	       $  & $7.26 \times 10^{-1}$   &  --                \\
$r_c$          &   $5.03 \times 10^{-1}$  & $1.58 \times 10^{-1}$   &  {\rm Mpc}         \\
$\beta_g$      &   $8.76 \times 10^{-1}$  & $5.72 \times 10^{-2}$   &  --	         \\
$r_g$  	       &   $5.07 \times 10^{-2}$  & $2.40 \times 10^{-3}$   &  {\rm Mpc}         \\
$Z_0$	       &   $1.35               $  & $1.34 \times 10^{-1}$   &  {\rm  solar}      \\
$\zeta$        &   $2.05 \times 10^{-1}$  & $1.60 \times 10^{-2}$   &  --	         \\
$T_0$	       &   $2.85               $  & $1.27 \times 10^{-1}$   &  {\rm keV}         \\
$T_1$          &   $6.55               $  & $6.20 \times 10^{-1}$   &  {\rm keV}         \\
$r_{\rm iso}$  &   $3.75 \times 10{-2} $  & $3.37 \times 10^{-3}$   &  {\rm Mpc}         \\ 
$\xi$          &   $3.94 \times 10^{-1}$  & $9.13 \times 10^{-2}$   &   --               \\
$r_{\rm tail}$ &   $1.26               $  & $5.80 \times 10^{-1}$   &  {\rm Mpc}         \\
\enddata
\tablecomments{The parameters not listed are  the column density of Galactic 
absorption $n_H = 1.21\times 10^{-22}$~cm$^{-2}$, 
$\varkappa$ frozen to 10, $r_Z$ frozen to  $8.24$~kpc i.e. 
one-half of the innermost  radial bin,  
$r_{\rm cool}$ (set equal to $r_c$) 
and $\omega = 0.2 \, \beta_c$.}

\end{deluxetable}

%
%
\begin{deluxetable}{llccc}
\tabletypesize{\small}
\tablecaption{\sc Checking the temperature distribution.\label{tab:temp}}
\tablehead{ 
	\colhead{Profile}                         & 
	\colhead{Free parameters}           &  	
	\colhead{$\chi^2$}                      &  
	\colhead{Degrees of freedom}      &
	\colhead{F-test statistic value}		   
}
\startdata
isothermal        &    $T_0$                                                                 &  4050.8   &     3035   & 82.7 \\ 
\hline
beta law           &   $T_0$  $r_{\rm tail}$  $\omega$                         &  3731.1   &    3033    & 32.7   \\
\hline
polytropic         &    $\dagger$                                                           &  3781.1    &    3034  &  35.6   \\
\hline
best-fit          &  $T_0$ $T_1$ $r_{\rm iso}$ $\xi$ $r_{\rm tail}$    &  3652.3   &    3031  &   --     \\
\enddata
\tablecomments{The number of free parameters required to describe correctly the temperature
distribution. In the beta-model the exponent is free. The F-test statistic values compare
the model of each line with respect to the best-fit one. In the row of the polytropic 
temperature model the names of the free parameters have not been reported, since the 
function (\ref{eq:kt}) has been modified. In this case, the temperature 3-D distribution
is described by two parameters.  In all  cases the associated probability
is well below $10^{-10}$, and has not been reported.}
\end{deluxetable}

\end{document}